\documentclass[final,5p,times,twocolumn]{elsarticle}

\usepackage{hyperref}
\usepackage{enumitem}
\usepackage{float}
\usepackage{amsmath}
\usepackage{makecell}
\usepackage{algorithm}
\usepackage{subfig}
\usepackage{stmaryrd}
\usepackage{xspace}
\usepackage{comment}
\usepackage[noend]{algpseudocode}
\newtheorem{theorem}{Theorem}

\newtheorem{proposition}{Proposition}
\newtheorem{lemma}{Lemma}

\newtheorem{example}{Example}

\newcommand{\tgds}{\textsf{tgds}\xspace}
\newcommand{\tgd}{\textsf{tgd}\xspace}
\newcommand{\egds}{\textsf{egds}\xspace}

\newcommand{\ctgds}{\textsf{ctgds}\xspace}

\journal{Journal of \LaTeX\ Templates}

\begin{document}

\begin{frontmatter}

\title{Reasoning on Property Graphs with Graph Generating Dependencies}

\author[1]{Larissa C. Shimomura}\corref{cor1}
\ead{l.capobianco.shimomura@tue.nl}
\author[1]{Nikolay Yakovets}
\ead{hush@tue.nl}
\author[1]{George Fletcher}
\ead{g.h.l.fletcher@tue.nl}

\cortext[cor1]{Corresponding author}

\address[1]{Eindhoven University of Technology, Department of Mathematics and Computer Science , PO Box 513 5600 MB - Eindhoven, The Netherlands}

\begin{abstract}
Graph Generating Dependencies (GGDs) informally express constraints between two (possibly different) graph patterns which enforce relationships on both graph's \emph{data} (via property value constraints) and its \emph{structure} (via topological constraints).
Graph Generating Dependencies (GGDs) can express tuple- and equality-generating dependencies on property graphs, both of which find broad application in graph data management.
In this paper, we discuss the reasoning behind GGDs. We propose algorithms to solve the \emph{satisfiability}, \emph{implication}, and \emph{validation} problems for GGDs and analyze their complexity.
To demonstrate the practical use of GGDs, we propose an algorithm which finds inconsistencies in data through validation of GGDs. Our experiments show that even though the validation of GGDs has high computational complexity, GGDs can be used to find data inconsistencies in a feasible execution time on both synthetic and real-world data.
\end{abstract}


\end{frontmatter}

\section{Introduction}

Constraints play a key role in data management research, e.g., in the study of data quality, data integration and exchange, and query optimization~\cite{Barcelo0R13,Bohannon2007,2012Fan,Fan19,Fan2019a,Fan2016,FrancisL17,IlyasC19}.
The use of graph-structured data sets has increased in different domains, such as social networks, biological networks and knowledge graphs. As consequence, the study of graph dependencies is also of increasing practical interest~\cite{Bonifati2018,Fan19} and it also raises new challenges as graphs are typically schemaless and relationships are first-class citizens.

To address this practical need, recently, different classes of dependencies for graphs have been proposed, for example, Graph Functional Dependencies (GFDs~\cite{Fan2016}), Graph Entity Dependencies (GEDs~\cite{Fan2019a}) and Graph Differential Dependencies (GDDs~\cite{Kwashie2019}). 
However, these types of dependencies focus on generalizing functional dependencies (i.e., variations of \emph{ equality}-generating dependencies) and cannot fully capture \emph{tuple}-generating dependencies (\tgds) for graph data~\cite{Fan19}.

As an example, we might want to enforce the constraint on a human resources graph that ``if two \emph{people} vertices have the same \emph{name} and \emph{address} property-values and they both have a {\em works-at} edge to the same \emph{company} vertex, then there should be a \emph{same-as} edge between the two people''.  This is an example of a \tgd on graph data, as satisfaction of this constraint requires the existence of an edge (i.e., the \emph{same-as} edge), and when it is not satisfied, the graph is \emph{repaired} by generating \emph{same-as} edges where necessary.
\tgds are important for many applications, e.g., data cleaning and integration~\cite{2012Fan,IlyasC19}.  

Indeed, \tgds arise naturally in graph data management applications. Given the lack of
\tgds for graphs in the current study of graph dependencies, a new
class of graph dependencies called Graph Generating Dependencies (GGDs) have been proposed~\cite{Shimomura2020}. The GGDs
fully support \tgds for property graphs (i.e., \tgds for graphs where vertices and edges can have associated property values, such as names and addresses in our example above -- a very common data model in practical graph data management systems) and generalize earlier graph
dependencies.  
The main contribution of the GGDs is the ability to extend \tgds to graph data. 
Informally, a GGD expresses a constraint between two (possibly different) graph patterns enforcing relationships between property (data) values and
a topological structure.  

In this paper, we study three reasoning problems on property graphs with GGDs: satisfiability, implication and validation. By studying these problems we can understand how GGDs behave in applications and what are their limitations.
In our experiments, we show how the validation algorithm can be used to identify data inconsistencies according to GGDs. Our results show scenarios in which GGDs can be used in practice.

\section{Related Work}
\label{sec:related}

We place GGDs in the context of relational and graph dependencies. In the following subsections, we present the main types of relational and graph dependencies proposed in the literature related to aspects proposed for GGDs.

\subsection{Relational Data Dependencies}

The classical Functional Dependencies (FDs) have been 
widely studied and extended for contemporary applications in data management.
Related to the GGDs, other classes of dependencies proposed for relational data are the Conditional Functional Dependencies (CFDs~\cite{Bohannon2007,2012Fan}) and the Differential Dependencies (DDs~\cite{Song2011}).
CFDs were proposed for data cleaning tasks where the main idea is to enforce an FD only for a set of tuples specified by a condition, unlike the original FDs in which the dependency holds for the whole relation~\cite{Fan2008}.

DDs extend the FDs by specifying looser constraints according to user-defined distance functions between attribute values~\cite{Song2011}. That is, given two tuples $t_1$ and $t_2$, if the distance $\delta$ between two tuple attributes $x$, $\delta\{t_1(x), t_2(x)\}$ agree on a specified difference/threshold then the distance between their attributes $y$, $\delta\{t_1(y), t_2(y)\}$ should also agree on a specified difference/threshold.
Since DDs are defined according to thresholds, previously proposed algorithms for FDs discovery cannot be applied, hence, new discovery algorithms were proposed for this class of dependencies~\cite{Kwashie2014,Kwashie2015,Song2014}.
Besides the new approaches of the discovery algorithms, Song et. al~\cite{Song2011} also addressed its difference from previous classes of dependencies and analyzed the consistency problem of a set of DDs, the implication problem,  and defined the minimal cover of a set of DDs. 

Tuple-generating dependencies (\tgds) are a well-known type of dependency with applications in data integration and data exchange~\cite{Beeri1984,Fagin2003}. 
Special cases of \tgds and its extensions also have a wide range of applications~\cite{Ma2014}, e.g., the inclusion dependencies (INDs). An IND states that all values of a certain attribute combination are also contained in the values of another attribute combination~\cite{Dursch2019}. Kruse et al.~\cite{Kruse2016} studied the discovery of INDs, more specifically, conditional inclusion dependencies (CINDs) for RDF data~\cite{Kruse2016}. In this work, we use their discovery algorithm for generating input GGDs for our experiments.

An example of an extension of \tgds are the constrained tuple-generating dependencies (\ctgds)~\cite{Maher1996} which add a condition (a constraint) on variables. The paper by Maher and Srivastava~\cite{Maher1996} proposes two Chase procedures to solve the implication problem for \ctgds and the conditions in which these procedures can be terminated early.

\subsection{Graph Dependencies}

Graph dependencies proposed for the property graphs include the graph  functional dependencies (GFDs), graph entity dependencies (GEDs), 
and graph differential dependencies (GDDs)~\cite{Fan2016,Fan2019a,Kwashie2019}.
GFDs are formally defined as a pair $(Q[\overline{x}], X \rightarrow Y)$ in which $Q[\overline{x}]$ is a graph pattern that defines a topological constraint while $X, Y$ are two sets of literals that define the property-value functional dependencies of the GFD. 
The property-value dependency is defined for the vertex attributes present in the graph pattern. 
GEDs~\cite{Fan2019a} subsume GFDs and can express FDs, GFDs, and equality-generating dependencies (\egds). Besides the property-value dependencies present in GFDs, GEDs also carry special literals to enable identification of vertices in the graph pattern.
GDDs extend GEDs by introducing distance functions instead of equality functions, similar to DDs for relational data but defined over a topological constraint expressed by a graph pattern. 

Similar to the definition of our proposed GGDs, Graph Repairing Rules (GRRs~\cite{Cheng2018}) were proposed to express automatic repairing semantics for graphs. The semantics of a GRR is: given a source graph pattern it should be repaired to a given target graph pattern.
The graph-pattern association rules (GPARs~\cite{Fan2015}) is a specific case of \tgds and has been applied to social media marketing. A GPAR is a constraint of the form $Q(x,y) \Rightarrow q(x,y)$ which states that if there exists an isomorphism from the graph pattern $Q(x,y)$ to a subgraph of the data graph, then an edge labeled $q$ between the vertices $x$ and $y$ is likely to hold.

Other types of dependencies recently proposed for property graphs include Temporal Graph Functional Dependencies (TGFDs)~\cite{Alipourlangouri2022} and Graph Probabilistic Dependencies (GPDs)~\cite{Zada2020}. The main idea of TGFDs is to extend GFDs by adding a time constraint. Semantically, a TGFD enforces a topological constraint and a dependency to hold within a time interval. 
GPDs extend GFDs by including a probability value $P$ in which the sets of literal $Y$ is defined by $X$.  
Even though GPDs also introduce more relaxed constraints, their approach has a different semantic to the use of differential/similarity constraints of GDDs and GGDs.

The main differences of the proposed GGDs compared to previous works are: (i) the use of differential constraints, (ii) edges are treated as first-class citizens in the graph patterns (in alignment with the property graph model) and, (iii) the ability to entail the generation of new vertices and edges (see Section ~\ref{sec:definition} for details). 
With these new features, GGDs can encode relations between two graph patterns as well as the (dis)similarity between its vertices and edges properties values. 
In general, GGD is the first constraint formalism for property graphs supporting both \egds and \tgds.

In the first study of GGDs~\cite{Shimomura2020}, the authors introduced the definition and semantics of the this new class of dependencies. In this paper, we focus on three fundamental reasoning problems of GGDs over property graphs and show how GGDs can be used in practice to detect data inconsistencies.

\section{Preliminaries}
\label{sec:preliminaries}

In this section, we summarize standard notation and concepts that will be used throughout the paper~\cite{Fan2019a,Song2011,Bonifati2018,Shimomura2020}.
Let $O$ be a set of objects, $L$ be a finite set of labels, $K$ be a set of property keys, and $N$ be a set of values. We assume these sets to be pairwise disjoint. 
A {\bf property graph} is a structure $(V,E,\eta, \lambda, \nu)$ where:
\begin{itemize}
    \item $V \subseteq O$ is a finite set of objects, called vertices;
    \item $E \subseteq O$ is a finite set of objects, called edges;
    \item $\eta: E \rightarrow V \times V$ is function assigning to each edge an ordered pair of vertices;
    \item $\lambda: V \cup E \rightarrow P(L)$ is a function assigning to each object a finite set of labels (i.e., $P(S)$ denotes the set of finite subsets of set $S$). 
    Abusing the notation, we will use $\lambda_v$ for the function assigning labels to vertices and $\lambda_e$ for the function that assigns labels to the edges; and
    \item $\nu: (V \cup E) \times K \rightarrow N$ is partial function assigning values for properties/attributes to objects, such that the object sets $V$ and $E$ are disjoint (i.e., $V \cap E = \emptyset$) and the set of domain values where $\nu$ is defined is finite.
\end{itemize}

A {\bf graph pattern} is a directed graph $Q[\overline{x}] = (V_Q, E_Q, \lambda_Q, \eta_Q)$ where $V_Q$ and $E_Q$ are finite sets of pattern vertices and edges, respectively, and $\lambda_Q$ is a function that assigns a label $\lambda_Q(u)$ to each vertex $u \in V_Q$ or edge $e \in E_Q$, and $\eta_Q$ assigns to each edge an ordered pair of vertices. 
Abusing notation, we use ${\lambda_v}_Q$ as a function to assign labels to vertices and ${\lambda_e}_Q$ to assign labels to edges. 
Additionally, $\overline{x}$ is a list of variables that include all the vertices in $V_Q$ and edges in $E_Q$. 

We say a label $l$ {\bf matches} a label $l' \in L$, denoted as $l \asymp l'$, if $l \in L$  and $l = l'$ or $l =$ `-' (wildcard) . 
A match denoted as $h[\overline{x}]$ of a graph pattern $Q[\overline{x}]$ in a graph G is a homomorphism of $Q[\overline{x}]$ to G such that for each vertex $u \in V_Q, {\lambda_v}_Q(u) \asymp {\lambda_v}(h(u))$; and for each edge $e = (u,u') \in E_Q$, there exists an edge $e' = (h(u), h(u'))$ and ${\lambda_e}_Q(e) \asymp {\lambda_e}(e')$.

We denote as $\llbracket Q[\overline{x}],\phi \rrbracket_G$ the evaluation results of the graph pattern query $Q[\overline{x}]$ on the graph G, such that the results of this evaluation also satisfy constraints in $\phi$, i.e., $Q[\overline{x}] \models \phi$.  

A {\bf differential function}  $\phi[A]$ on attribute $A$ is a constraint of difference over $A$ according to a distance metric~\cite{Song2011}. Given two tuples $t_1,t_2$ in an instance I of relation R, $\phi[A]$ is true if the difference between $t_1.A$ and $t_2.A$ agrees with the constraint specified by $\phi[A]$, where $t_1.A$ and $t_2.A$ refers to the value of attribute $A$ in tuples $t_1$ and $t_2$, respectively. We use the differential constraint idea to define constraints over attributes in GGDs.

\section{GGD: Syntax and Semantics}
\label{sec:definition}

In this section, we review the syntax and semantics of the GGDs in~\cite{Shimomura2020}.
A {\bf Graph Generating Dependency} (GGD) is a dependency of the form \[ Q_s[\overline{x}], \phi_s \rightarrow Q_t[\overline{x},\overline{y}],\phi_t\] where:

\begin{itemize}
    \item $Q_s[\overline{x}]$ and $Q_t[\overline{x},\overline{y}]$ are graph patterns, called \textbf{source} graph pattern and \textbf{target} graph pattern, respectively;
    \item  $\phi_s$ is a set of differential constraints defined over the variables $\overline{x}$ (variables of the graph pattern $Q_s$); and
    \item $\phi_t$ is a set of differential constraints defined over the variables $\overline{x} \cup \overline{y}$, in which $\overline{x}$ are the variables of the source graph pattern $Q_s$ and $\overline{y}$ are any additional variables of the target graph pattern $Q_t$.
\end{itemize}

A differential constraint in $\phi_s$ on $[\overline{x}]$ (resp., in $\phi_t$ on $[\overline{x},\overline{y}]$) is a constraint of one of the following forms~\cite{Kwashie2019,Song2011}:
\begin{enumerate}
    \item $\delta_A(x.A,c) \le t_A$ 
    \item $\delta_{A_1A_2}(x.A_1, x'.A_2) \le t_{A_1A_2}$
    \item $x = x'$ or $x \neq x'$
\end{enumerate}
where $x, x' \in \overline{x}$ (resp. $\in \overline{x} \cup \overline{y}$) for $Q_s[\overline{x}]$ (resp. for $Q_t[\overline{x},\overline{y}]$), $\delta_A$ is a user-defined \emph{similarity function} for the property $A$ and $x.A$ is the property value of variable $x$ on $A$, $c$ is a constant of the domain of property $A$ and $t_A$ is a pre-defined threshold. 
The differential constraints defined by (1) and (2) can use the operators $(=, <, >, \le, \ge, \neq)$. 

The constraint (3) $x = x'$ states that $x$ and $x'$ are the same entity (vertex/edge) and it can also use the inequality operator stating that $ x \neq x'$. An important feature of GGDs is that both vertices and edges are considered variables (in source and target graph patterns), which allows to compare vertex-vertex variables, edge-edge and vertex-edge variables. 

Consider a graph pattern $Q[\overline{z}]$, a set of differential constraints $\phi_z$ and a match of this pattern represented by $h[\overline{z}]$ in a graph $G$. 
The match $h[\overline{z}]$ satisfies $\phi_z$, denoted as $h[\overline{z}] \models \phi_z$ if the match $h[\overline{z}]$ satisfies every differential constraint in $\phi_z$. If $\phi_z = {\emptyset}$ then $h[\overline{z}] \models \phi_z$ for any match of the graph pattern $Q[\overline{z}]$ in $G$.

Given a GGD $Q_s[\overline{x}], \phi_s \rightarrow Q_t[\overline{x},\overline{y}],\phi_t$ we denote the matches of the source graph pattern $Q_s[\overline{x}]$ as $h_s[\overline{x}]$ while the matches of the target graph pattern $Q_t[\overline{x},\overline{y}]$ are denoted by $h_t[\overline{x},\overline{y}]$ which can include the variables from the source graph pattern $\overline{x}$ and additional variables $\overline{y}$ particular to the target graph pattern.  

A GGD $\sigma = Q_s[\overline{x}], \phi_s \rightarrow Q_t[\overline{x},\overline{y}],\phi_t$ holds in a graph G, denoted as $G \models \sigma$, if and only if for every match $h_s[\overline{x}]$ of the source graph pattern $Q_s[\overline{x}]$ in $G$ satisfying the set of constraints $\phi_s$, there exists a match $h_t[\overline{x},\overline{y}]$ of the graph pattern $Q_t[\overline{x},\overline{y}]$ in $G$ satisfying $\phi_t$ such that for each $x$ in $\overline{x}$ it holds that $h_s(x) = h_t(x)$. 
If a GGD does not hold in $G$ then it is \emph{violated}. Such violations can be fixed by \emph{generating} new vertices/edges in graph $G$ according to the violated GGD.

\section{Properties of GGDs}
\label{sec:properties}

In this section, we define properties and concepts of GGDs that are used in the proposed algorithms to solve the reasoning problems we will study in Section~\ref{sec:reasoning}. 

\paragraph{\textbf{Subjugation}} To define subjugation for GGDs, we adapt the subjugation definition on the sets of differential constraints from the GDDs~\cite{Kwashie2019} and subsumption of DDs~\cite{Song2011}.  
A differential function $\phi_1$ subjugates a second differential function $\phi_2$ if for any graph pattern match $h[\overline{x}]$, if $h[\overline{x}]$ satisfies $\phi_2$ ($h[\overline{x}] \models \phi_2$), then $h[\overline{x}]$ also satisfies $\phi_1$ ($h[\overline{x}] \models \phi_1$)~\cite{Song2011}.

A set of GGD differential constraints $\tau$ subjugates another set of GGD differential constraints $\omega$, denoted as $\tau \succeq \omega$, iff: 
\begin{enumerate}
    \item for every constraint $\delta_{A}(x.A,c) \leq t_{A} \in \tau$, there exists $\delta_{A}(x.A, c) \leq t_{A}' \in \omega$ and $t_{A} \ge t_{A}'$;
    \item for every constraint $\delta_{A_1A_2}(x.A_1,x'.A_2) \leq t_{A_1A_2} \in \tau$, there exists $\\ \delta_{A_1A_2}(x.A_1, x'.A_2) \leq t_{A_1A_2}' \in \omega$ and $t_{A_1A_2} \ge t_{A_1A_2}'$;
    \item for every constraint $x = x' \in \tau$ there exists $x = x' \in \omega$; 
\end{enumerate}

Given the transitivity and properties of differential constraints studied by~\cite{Song2011}, we have the following proposition.

\begin{proposition}
    Suppose $\tau, \omega$ and $\psi$ are sets of differential constraints in GGDs, if $\tau \succeq \psi$ and $\psi \succeq \omega$ then $\tau \succeq \omega$~\cite{Song2011}. 
\end{proposition}

\paragraph{\textbf{Graph query containment}}
Graph pattern queries can be expressed as conjunctive queries (CQs)~\cite{Bonifati2018}.
According to Chandra and Merlin~\cite{Chandra1977}, given two conjunctive queries $q_1,q_2$, $q_1 \sqsubseteq q_2$ is equivalent to the existence of a homomorphism $h$ from $q_2$ to $q_1$. 
Given this definition, we have the following proposition.

\begin{proposition}
    A graph pattern query $Q_1$ is contained in a graph pattern query $Q_2$ denoted as $Q_1 \sqsubseteq Q_2$, if the answer set of $Q_1$ is contained in the answer set of $Q_2$ for every possible property graph $G$, which means that there exists a homomorphism $h$ from $Q_2$ to $Q_1$.
\end{proposition}

    \begin{figure}[htb]
  \centering
  \includegraphics[width=0.8\linewidth]{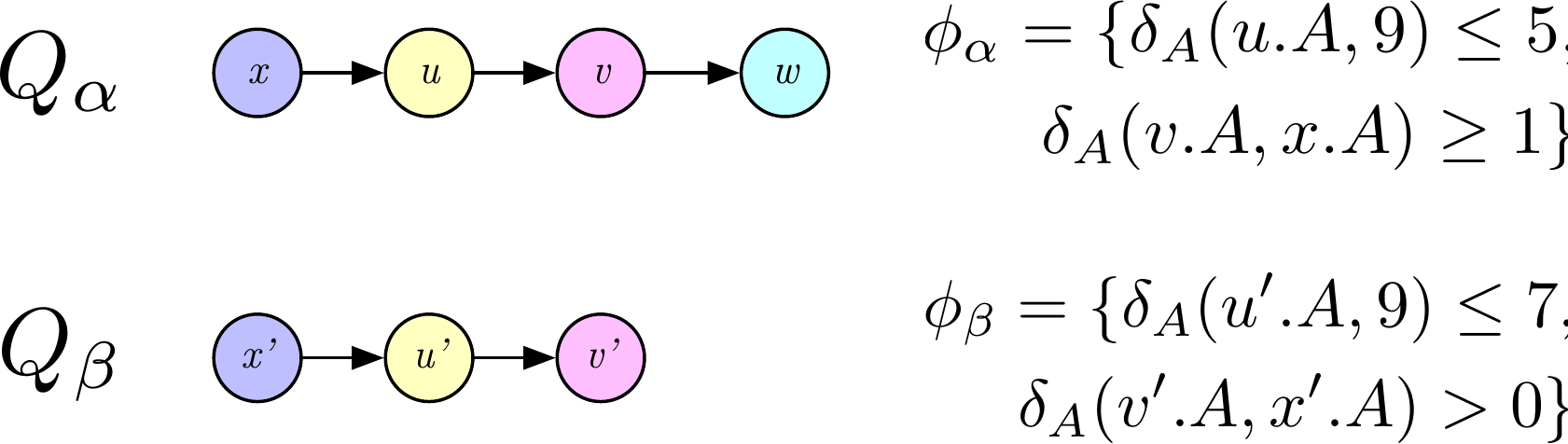}
  \caption{Containment Example}
  \label{fig:exampleContaiment}
\end{figure}

\begin{example}
Consider the graph patterns $Q_\alpha$ and $Q_\beta$ and its respective sets of differential constraints $\phi_\alpha$ and $\phi_\beta$ in \autoref{fig:exampleContaiment}. Observe that exists an homomorphic mapping of the nodes $h(x')= x, h(u')= u, h(v')= v$ in $Q_\alpha$ to the nodes in $Q_\beta$, therefore we can conclude that $Q_{\alpha} \sqsubseteq Q_{\beta}$. Considering this homomorphic mapping from $Q_\alpha$ to $Q_\beta$, the differential constraint of $\phi_\beta$, $\delta_A(u'.A, 9) \le 7$ subjugates the differential constraint $\delta_A(u.A,9) \le 5$ of $\phi_\alpha$ and the differential constraint of $\phi_\beta$, $\delta_A(v'.A, x'.A) > 0$ subjugates the differential constraint $\delta_A(v.A,x.A) \ge 1$ of $\phi_\alpha$ as any value that satisfies the differential constraints of $\phi_\alpha$ also satisfies the differential constraints of $\phi_\beta$. Consequently, we can conclude that $\phi_\beta \succeq \phi_\alpha$. 
\end{example}

\paragraph{\textbf{Infeasibility and disjoint differential constraints}}

A set of constraints $\phi$ is infeasible if there does not exist a nonempty graph pattern match $h[\overline{x}]$ of a graph pattern $Q[\overline{x}]$ over any nonempty graph $G$, such that $h[\overline{x}] \models \phi$.
Two differential constraints $\phi_a$ and $\phi_b$ are disjoint if their intersection (conjunction) is infeasible ($\phi_a \wedge \phi_b$=infeasible). The conjunction of a differential constraint $\phi_a$ with its complement $\overline{\phi_a}$ ($\phi_a \wedge \overline{\phi_a}$), is always infeasible~\cite{Song2011}.

\begin{figure}[t]
  \centering
  \includegraphics[width=0.9\linewidth]{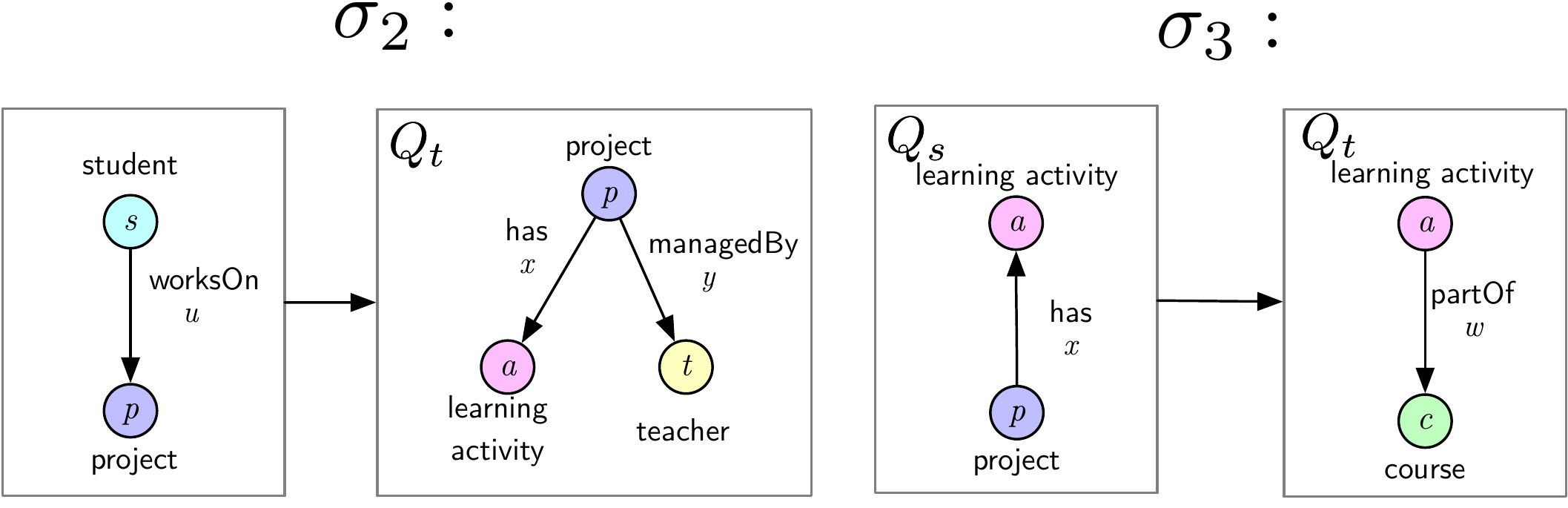}
  \caption{Example GGDs}
  \label{fig:transGGDs}
\end{figure}

\begin{example}
Consider the GGD $\sigma_2$ of \autoref{fig:transGGDs} and the set of differential constraints $\phi = \{\delta_{\text{hours}}(\text{s.hours}, 10) > 5, \delta_{\text{hours}}(\text{s.hours}, 10) < 5 \}$, the set $\phi$ is infeasible as the differential constraints  $\delta_{\text{hours}}(\text{s.hours}, 10) > 5$ and $\delta_{\text{hours}}(\text{s.hours}, 10) < 5$ are disjoint. In conclusion, there are no nonempty matches of the graph pattern that can satisfy the set $\phi$.
\end{example}  

\paragraph{\textbf{GGDs interaction}} A GGD $\sigma_1 = Q_{s1}[\overline{x}],\phi_{s1} \rightarrow Q_{t1}[\overline{x},\overline{y}],\phi_{t1}$ interacts on the source (resp. on the target) with a GGD $\sigma_2 = Q_{s2}[\overline{x}],\phi_{s2} \rightarrow Q_{t2}[\overline{x},\overline{y}],\phi_{t2}$ if and only if the intersection between $Q_{s1}[\overline{x}]\phi_{s1}$ and $Q_{s2}[\overline{x}]\phi_{s2}$ (resp. $Q_{t1}[\overline{x},\overline{y}]\phi_{t1}$ and $Q_{t2}[\overline{x},\overline{y}]\phi_{t2}$) is not empty.
Considering $Q_{s1}=\{V_{s1}, E_{s1}, \lambda_{s1}, \eta_{s1}\}$ and $Q_{s2} = \{V_{s2}, E_{s2}, \lambda_{s2}, \eta_{s2}\}$, we define the intersection of  $Q_{s1}[\overline{x}]\phi_{s1}$ and $Q_{s2}[\overline{x}]\phi_{s2}$ as a graph pattern $Q_\cap = \{V_\cap, E_\cap, \lambda_\cap, \eta_\cap\}$ in which $Q_\cap$ is the maximal graph pattern (maximal set of subgoals of a conjunctive query) of $Q_{s1}$ and $Q_{s2}$ such that there exists a homomorphic mapping from $Q_\cap$ to $Q_{s1}$ and from $Q_\cap$ to $Q_{s2}$. We define $Q_\cap$ as:
\begin{itemize}
    \item $V_\cap$ is the set of nodes $v \in V_{s1}$ such that there exists $\lambda_{s1}(v) \asymp \lambda_{s2}(v')$ or $\lambda_{s2}(v') \asymp \lambda_{s1}(v)$, where $v' \in V_{s2}$. Additionally, the differential constraints in $\phi_{s1}$ that refer to $v \in V_{s1}$ are feasible to the differential constraints in $\phi_{s2}$ that refer to $v' \in V_{s2}$.
    \item $E_\cap$ is the set of edges $e \in E_{s1}$ such that there exists $\lambda_{s1}(e) \asymp \lambda_{s2}(e')$ or $\lambda_{s2}(e') \asymp \lambda_{s1}(e)$, where $e' \in E_{s2}$.
    The labels of the source and target of $e$ and $e'$ also match, formally, considering $\eta_{s1}(e) = (v_1,v_2)$ and $\eta_{s2} = (v_3,v_4)$, $\lambda_{s1}(v_1) \asymp \lambda_{s2}(v_3)$ and $\lambda_{s1}(v_2) \asymp \lambda_{s2}(v4)$ are true.
    Additionally, the differential constraints in $\phi_{s1}$ that refer to $e \in E_{s1}$ are feasible to the differential constraints in $\phi_{s2}$ that refer to $e' \in E_{s2}$.
    \item $\lambda_\cap$ is the function $\lambda_{s1}$ which assigns the labels to the nodes and edges in $Q_\cap$ (in this case, same labels as in $Q_{s1}[\overline{x}]$).
    \item $\eta_\cap$ is the function $\eta_{s1}$ which assigns a pair of vertices in $V_\cap$ to $E_\cap$.
\end{itemize}
The graph pattern $Q_\cap$ is not empty if $V_\cap \neq \{\emptyset\}$ or $E_\cap \neq \{\emptyset\}$.
Informally, two GGDs interact if their sources or targets can possibly match some of the same nodes and/or edges in a graph $G$. 

\begin{example}
Consider the GGDs in \autoref{fig:transGGDs} with empty $\phi_s$ and $\phi_t$. In this example, the GGD $\sigma_2$ interacts on the source with the GGD $\sigma_3$. The GGD $\sigma_2$ and the GGD $\sigma_3$ can both match the same nodes in a graph $G$ to the node project $p$ present in the source of both GGDs.
\end{example} 

\paragraph{\textbf{Transitive GGDs}} Given a set of GGDs $\Sigma$, we say that a GGD $\sigma_1 \in \Sigma$ is transitive if there exists a GGD $\sigma_2 \in \Sigma$ in which the target side of $\sigma_2$ interacts with the source side of $\sigma_1$. Formally, consider $\sigma_1 = Q_{s1}[\overline{x}],\phi_{s1} \rightarrow Q_{t1}[\overline{x},\overline{y}],\phi_{t1}$ and $\sigma_2 = Q_{s2}[\overline{x}],\phi_{s2} \rightarrow Q_{t2}[\overline{x},\overline{y}],\phi_{t2}$, $\sigma_1$ is transitive if $Q_{t2}[\overline{x},\overline{y}],\phi_{t2} \Leftarrow Q_{s1}[\overline{x}],\phi_{s1}$ ($\llbracket Q_{t2}[\overline{x},\overline{y}],\phi_{t2} \rrbracket_G  \subseteq \llbracket Q_{s1}[\overline{x}],\phi_{s1} \rrbracket_G $).
Informally, $\sigma_1$ is transitive in $\Sigma$ if there exists a $\sigma_2$ in which its target side can trigger the source side of $\sigma_1$

\begin{example}\label{example:transitiveGGDs}
Consider the GGDs in \autoref{fig:transGGDs} with empty $\phi_s$ and $\phi_t$. In this example, we can clearly see that the GGD $\sigma_3$ is transitive as what is enforced by the target of $\sigma_2$ interacts with the source side of $\sigma_3$. When repairing a graph $G$ according to $\sigma_2$ we can possibly generate new matches of the source graph pattern of $\sigma_3$. Which means that the repairing of a GGD can also trigger the repairing of another GGD. 
\end{example}

\section{The Chase Procedure for GGDs}
\label{sec:chase}

The Chase procedure was originally proposed for testing logical implication between sets of dependencies~\cite{Beeri1984} and has gained attention due to its application in data exchange, query optimization and data repair~\cite{IlyasC19}. 
In this section, we define the Chase procedure for GGDs to solve the satisfiability and implication problems. Our Chase procedure is based on the standard Chase procedure for \tgds and GEDs~\cite{Maher1996,Fan2019a}.

\begin{figure*}[t]
  \centering
  \includegraphics[width=0.7\linewidth]{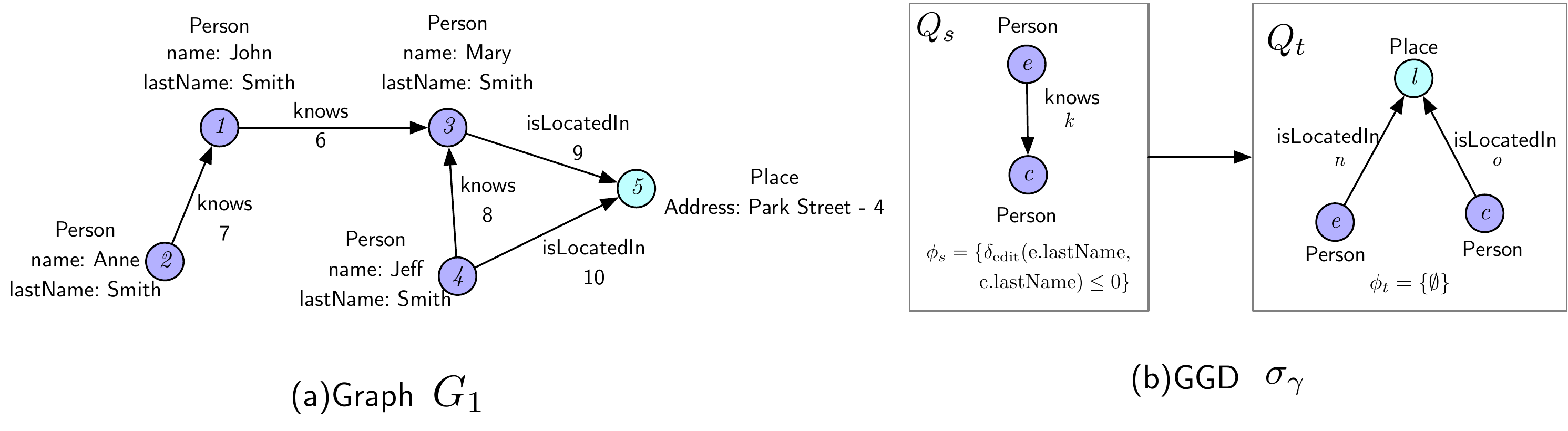}
  \caption{Chase Example}
  \label{fig:chase1}
\end{figure*}

The Chase procedure for \ctgds and GEDs were both defined considering the equality of attributes. Given the differential constraints in $\phi_s$ and $\phi_t$ of the GGDs, we use the idea of range of values in this Chase procedure.
We use the graph $G_1$ and the GGD $\sigma_\gamma$ in \autoref{fig:chase1} as a running example in this section. 

\subsection{The Chase for GGDs}

Following the Chase methods designed for GEDs~\cite{Fan2019a} and \tgds~\cite{greco2012}, we define the Chase method according to range relations.
Consider a graph $G = (V,E,\eta, \lambda, \nu)$ and a set $\Sigma$ of GGDs, we define the Chase for GGDs as a sequence of Chase steps on $G$. Next, we define the concept of range classes and Chase steps.

\paragraph{Range Class} For each $x \in (V \cup E)$, its range class (denoted as $[x]_{R_q}$) is a set of nodes or edges $y \in (V \cup E)$ of $G$ that are identified as $x$. The node or edge $y$ can be identified as $x$ because of the enforced constraints on GGDs of the type $x = y$ (see GGDs definition on Section~\ref{sec:definition}). 
For each attribute $A$ of $x \in (V \cup E)$ formally denoted as $\nu(x,A)$ , its range class is denoted as $[x.A]_{R_q}$. Informally, the range class refers to the possible set of values that an attribute can have enforced by a GGD in $\Sigma$. 

According to a GGD, the differential constraints defines the possible values of an attribute, as a consequence, we use the same semantics of the differential constraints to define the attribute range class.
For each attribute $A$ of $x \in (V \cup E)$, its range class $[x.A]_{R_q}$ contains a set of range class quadruples named $rcq = (\delta_A, val_A, t_A, op_A)$ that refers to the differential constraints of a GGD involving $x.A$, in which:
\begin{itemize}
    \item $\delta_A$ is the distance used in the differential constraint;
    \item $val_A$ is the set of attribute values that $\nu(x.A)$ is compared to;
    \item $t_A$ is the threshold used in the differential constraint and,
    \item $op_A$ is the operator of the differential constraint that is being compared to $(<, >, \leq, \geq, =, \neq)$
\end{itemize}
A $rcq_1 = (\delta_A, val_A, t_A, op_A)$ subsumes a second $rcq_2 = (\delta_B, val_B, t_B, op_B)$ according to the same subjugation properties of a differential constraint (see Section~\ref{sec:properties}). A range class $[x.A]_{R_q}$ subsumes a range class $[y.B]_{R_q}$, denoted as $[x.A]_{R_q} \succeq [y.B]_{R_q}$, if (i)$x$ and $y$ are the same node or edge on $G$ and, (ii) every $rcq$ $(\delta_B, val_B, t_B, op_B) \in [y.B]_{R_q}$ is subsumed by a $rcq$ $(\delta_A, val_A, t_A, op_A) \in [x.A]_{R_q}$.

\paragraph{Chasing - Initialization} - Given a graph $G = (V, E,\eta, \lambda, \nu)$ and a set $\Sigma$ of GGDs, we start the Chase procedure by initializing the range classes of each node and edge of $G$. For each $x \in (V \cup E)$, $[x]_{R_q}$ = $\{x\}$ and each attribute range class $[x.A]_{R_q} = \{(,x.A, 0, =)\}$, that means initializing each range class with a $rcq$ that indicates its own value.

\begin{figure}[t]
  \centering
  \includegraphics[width=\linewidth]{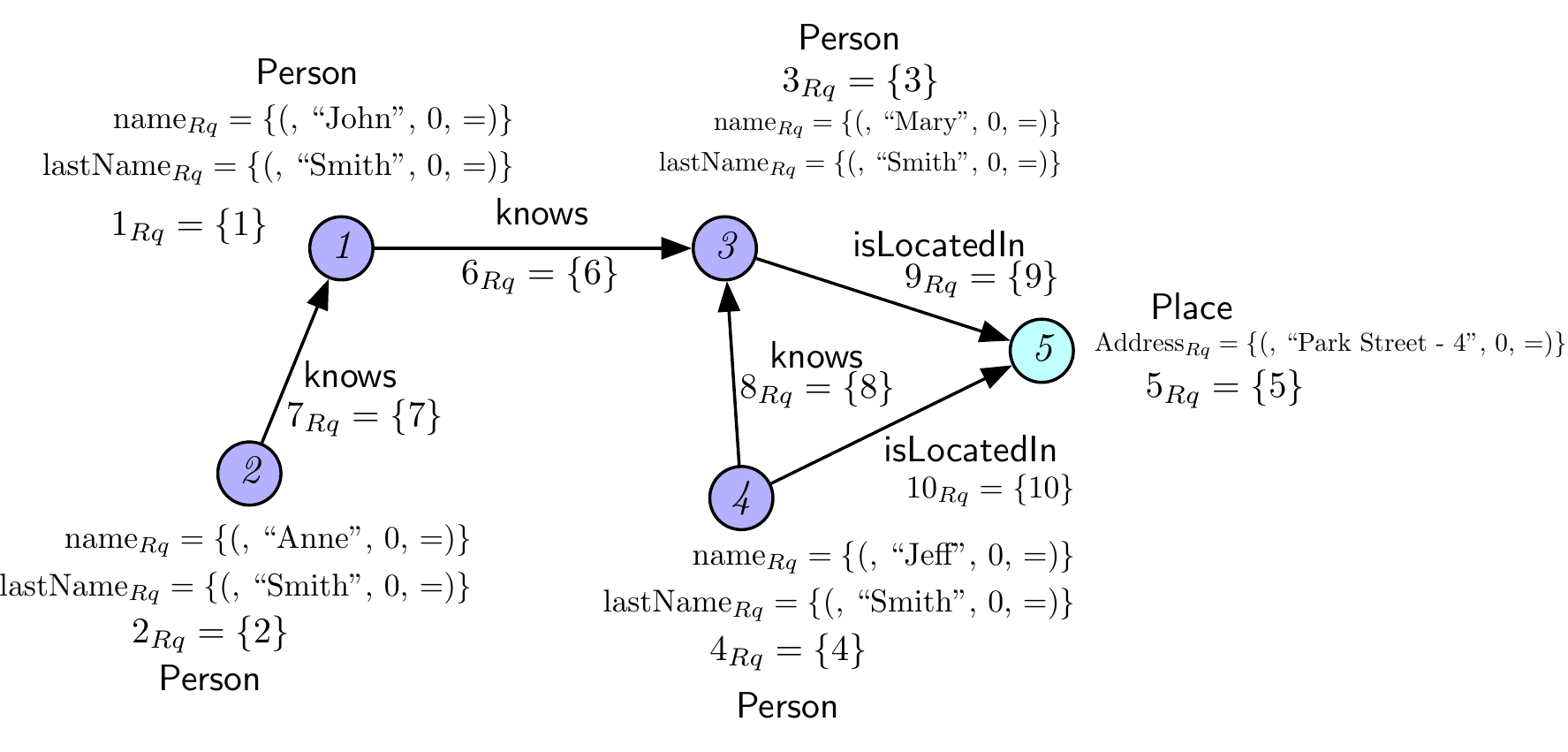}
  \caption{Range Classes Initialization}
  \label{fig:chase2}
\end{figure}

\begin{example}
To run Chase over the graph $G_1$ in \autoref{fig:chase1}, we initialize the range classes of each node and vertex of $G$ as well as the range classes of its attributes (see \autoref{fig:chase2}). As mentioned, we initialize each range class with an $rcq$ that indicates its own value. Observe that, in this case, each $rcq$ is initialized without a distance function attached, to represent the attribute value. In case this attribute was built according to a differential constraint, the $rcq$ would be initialized according to this differential constraint.
\end{example}

\paragraph{Chasing - Match} - After initializing the range classes of each node and edge in $G$, to apply each Chase step according to a GGD $\sigma \in \Sigma$, we need to find matches of the source side and the target side of $\sigma$ in $G$. In the Chase, a match $h_s[\overline{x}]$ of a graph pattern $Q[\overline{x}]$ is a homomorphism of $Q[\overline{x}]$ to G such that for each vertex $u \in V_Q, {\lambda_v}_Q(u) \asymp {\lambda_v}(h(u))$; and for each edge $e = (u,u') \in E_Q$, there exists an edge $e' = (h(u), h(u'))$ and ${\lambda_e}_Q(e) \asymp {\lambda_e}(e')$. And, we check if $h_s[\overline{x}] \models \phi_s$ according to the information in the range classes of each vertex and edge of $h_s[\overline{x}]$. Formally, for each vertex and edge $u$ in $h_s[\overline{x}] \models \phi_s$, if its range class $[u]_{R_q}$ subsumes the possible values of $u$ constrained by $\phi_s$ and its attributes range classes $[u.A]_{R_q}$ subsumes the possible values of $u.A$ constrained by $\phi_s$. 

\paragraph{Chasing - Step}
The Chase procedure is a sequence of Chase steps denoted as:
\[G \xrightarrow{(\sigma, h_s[\overline{x}] \models \phi_s)} G'\]

A Chase step according to a GGD $\sigma \in \Sigma$ can be applied if there exists a match of its source side, $h_s[\overline{x}] \models \phi_s$ in $G$. 
Each Chase step expands $G$ by generating new nodes or edges and performs a set of updates on the range classes of nodes and edges according to $\sigma$ and a match $h_s[\overline{x}]$ of the source graph pattern of $\sigma$, the result of these updates is a graph $G'$. These updates are executed according to the following steps:
\begin{enumerate}
    \item Update the range classes of the nodes and edges in $h_s[\overline{x}]$ according to $\phi_s$ (see \nameref{rangeUpdates});
    \item Search for a match of the target side of the GGD $\sigma$: (1) If there are no matches of the target side, we add new nodes and edges to $G$ and initialize its range classes according to $\phi_t$. (2) If there are matches of the target side, for each match we update its range classes according to $\phi_t$ (see \nameref{generationUpdates}). 
    \item Check if the Chase step is consistent. If the step is not consistent then stop the Chase (see \nameref{consistencyUpdates}).
\end{enumerate}
Next, we give details on each one of the mentioned updates.

\paragraph{Range Classes Updates - $\phi_s$ and $\phi_t$}\label{rangeUpdates}
Given $h_s(\overline{x})$, for each node or edge $x$ matched in $h_s[\overline{x}]$, get the differential constraints in $\phi_s$ that refer to $x$, denoted as $\phi_x$. 
\begin{itemize}
    \item Consider that $x'$ is the variable names of $x$. If a constraint of the type $\delta(x'.A, c) \le t_A$ or $\delta(x'.A,y.B) \le t_{AB}$ is in $\phi_x$, rewrite it as a $rcq$ of the type $\Phi = (\delta, val, t, op)$. If exists a $rcq$ $(\delta_A, val_A, t_A, op_A)$ in $[x.A]_{R_q}$ that is subsumed by $\Phi$, we update the $rcq (\delta_A, val_A, t_A, op_A) \in [x.A]_{R_q}$ with the values of $\Phi$. Else, if such subsumption does not exist, add $\Phi$ to $[x.A]_{R_q}$ ($[x.A]_{R_q} = [x.A]_{R_q} \cup \{\Phi\}$).
    \item Consider that $x'$ and $y'$ are the variable names of $x$ and $y$. If a constraint of the type $x' = y'$ is in $\phi_x$, in which $y$ is a node or edge in $h_s(\overline{x})$. Add $y$ to the range class of $x$, $[x]_{R_q} = [x]_{R_q} \cup y$. 
\end{itemize}
Informally, we update each node or edge range class with the loosest threshold from $\phi_s$ hence each attribute has the largest number of possible values.

In case there exists a target match, $h_t[\overline{x,y}] \models \phi_t$, we enforce the constraints in $\phi_t$ in $G$ by updating the range classes of the nodes and edges that were matched in $h_t[\overline{x,y}]$. We update the range classes in the same way as we do with $\phi_s$ on $h_s[\overline{x}]$.

\paragraph{Generation of new nodes and edges}\label{generationUpdates}
Given the Chase step $(\sigma, h_s[\overline{x}] \models \sigma)$, we generate new nodes and edges in the graph $G$ if there are no matches of the target side $Q_t[\overline{x,y}]\phi_t$. We generate nodes and edges that refer to the set $\overline{y}$ of variables in $Q_t[\overline{x},\overline{y}]$ of the target graph pattern, as the set $\overline{x}$ refers to the source variables and exist in the graph (match $h_s[\overline{x}]$). For each variable $y \in \overline{y}$, if $y$ refers to an node in $Q_t[\overline{x,y}]$, we create a new node or edge $u$ in which: (i)$\lambda_{Q_t}(u) = \lambda(u)$ ($\lambda$ is a label assigning functions of the target graph pattern and for the graph G respectively), 
(ii) initialize its range class $[u]_{R_q} = \{u\}$ and if there exist a differential constraint in $\phi_t$ that refers to an attribute of $u$, suppose $u.C$, we create a range class $[u.C]_{R_q}$ and initialize it with a $rcq$ with the information of this differential constraint.
Finally, we add $u$ to the graph $G$, in case $u$ is a node $V = V \cup {u}$, or in case it is an edge $E = E \cup {u}$.

\paragraph{Consistency}\label{consistencyUpdates} - We say that a Chase step is not consistent in $G$ if: 
\begin{enumerate}
    \item there exists a node or an edge $y$ in $[x]_{R_q}$ in which the  $\lambda(x) \not\asymp \lambda(y)$ and $\lambda(y) \not\asymp \lambda(x)$ (label conflict) or,
    \item there exists a $rcq$ $\omega = (\delta_B, val_B, t_B, op_B)$ on an attribute $y.B$ in $[x.A]_{R_q}$ in which 
    $\omega \wedge [x.A]_{R_q}$ = not feasible.
\end{enumerate}
Otherwise, the Chase step is consistent.
When the Chase step is consistent, it means that there are no contradictions and the target side of the GGD was correctly enforced. By the end of the step, we can merge each node and each edge in $[x]_{R_q}$ and also its range classes $[x.A]_{R_q}$.

\paragraph{Chase - Termination} - The Chase procedure terminates if and only if one of the following conditions holds: 
\begin{enumerate}
    \item There are no possible updates/expansions on $G$ according to any $\sigma \in \Sigma$. In this case, the Chase terminates and the Chase steps sequence is valid.
    \item A Chase step is inconsistent. In this case, the Chase terminates and the Chase steps sequence is invalid. 
\end{enumerate}

\begin{figure}[t]
  \centering
  \includegraphics[width=\linewidth]{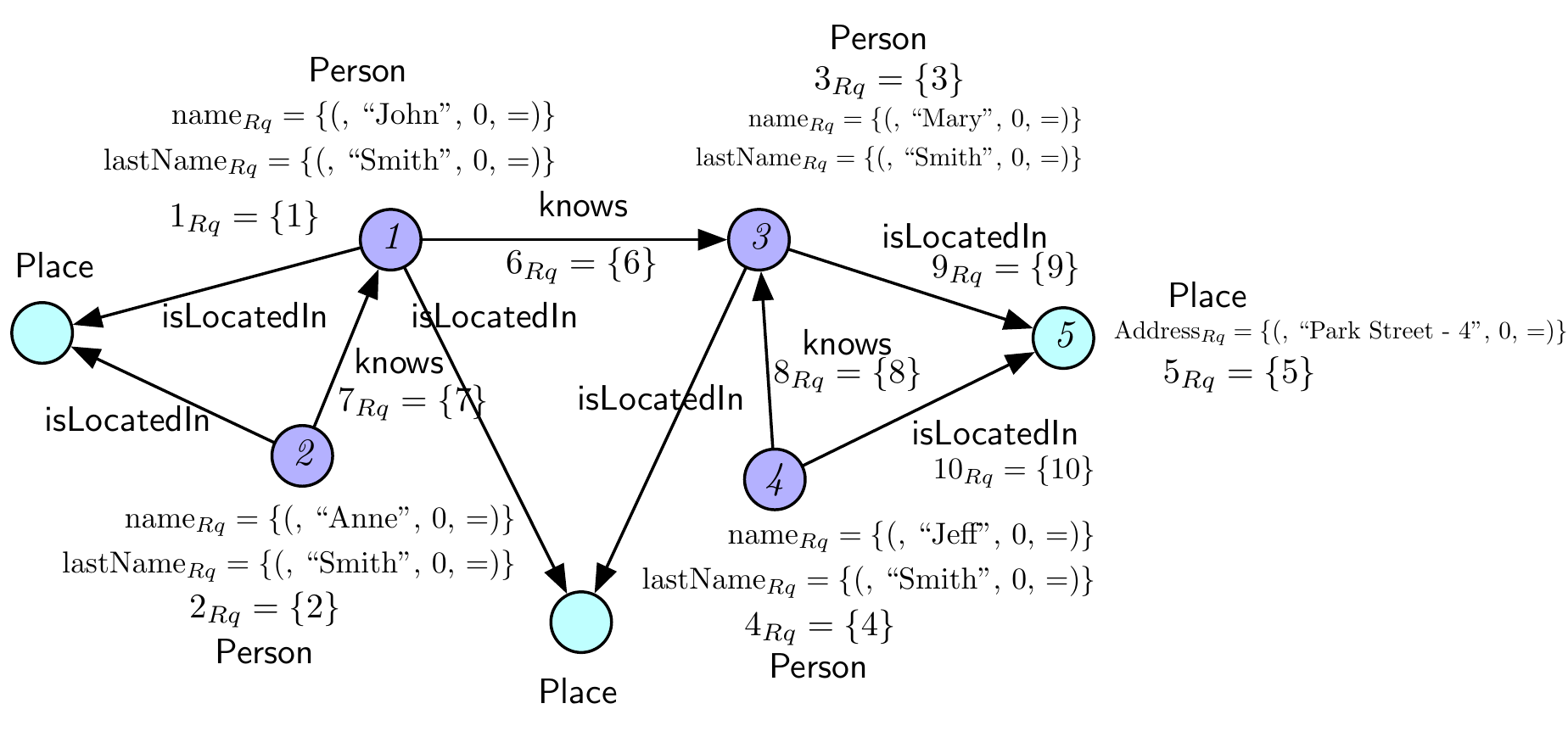}
  \caption{Chase Application of $\sigma_\gamma$}
  \label{fig:chase3}
\end{figure}

\begin{example}
\autoref{fig:chase3} shows the final results of the graph $G_1$ after the application of $\sigma_\gamma$. Observe that, to enforce the GGD $\sigma_\gamma$, new nodes of the label ``Place" and new edges of the label ``isLocatedIn" were generated. Since in $\sigma_\gamma$, the target differential constraints $\phi_t =\emptyset$, then the nodes and edges were generated without attributes. 
\end{example}

\paragraph{The Problem of Chase Termination}\label{para:chaseTermination} Analogous to \tgds and \ctgds, given the generating property of the GGDs, the Chase might not terminate even when there are no inconsistent Chase steps~\cite{Maher1996}.
Termination of different versions of the Chase procedure for \tgds have been long studied in the literature ~\cite{Calautti2015,Gogacz2020,greco2012}. In general, the Chase termination problem for \tgds is not decidable but there are positive results on sets of \tgds with specific syntactic properties~\cite{Calautti2015}.

A well-known property that has been proven that the Chase procedure terminates is when the set of \tgds is weakly-acyclic. 
Informally, a set of \tgds is weakly-acyclic if there is no cascading of generating null values during the Chase procedure~\cite{Fagin2005}. 
The problem of checking if a set of \tgds is weakly-acyclic is polynomial in the size of the set~\cite{greco2012} as it is verified by building a directed graph called dependency graph derived from the set of \tgds (see details in~\cite{Fagin2005}).
Besides weak-acyclicity, other generalizations and properties of sets of \tgds that can ensure Chase termination have been identified in the literature (see~\cite{greco2012} for a survey). 

\paragraph{Use of Chase for Repairing Graphs with GGDs} The Chase procedure is often used in the literature as a method to repair data according to a set of dependencies~\cite{Fan19,IlyasC19,Bohannon2007}. The repairing problem for the GGDs can also be seen as a method to enforce the target constraints of each GGD in a graph, which is the goal of the proposed Chase for GGDs. However, the Chase will always generate new nodes and edges which is a naive solution and might not always generate useful information. Therefore, new strategies should be proposed to choose when to generate and when to modify the existing graph. The problem of graph data repair using GGDs will be addressed in future work.

\section{Reasoning for GGDs}
\label{sec:reasoning}

In this section, we discuss the three following fundamental problems and their complexity:
\begin{itemize}
    \item \emph{Validation} - Given a set of GGDs $\Sigma$ and a non-empty graph $G$, does the set of GGDs $\Sigma$ hold in $G$, denoted as $G \models \Sigma$?
    \item \emph{Satisfiability} - A set of GGDs $\Sigma$ is satisfiable if (i) there exists a graph $G$ which is a model of $\Sigma$ ($G \models \Sigma$) and (ii) for each GGD $\sigma \in \Sigma$ there exists a match of $Q_s[\overline{x}]$ in $G$.  Given a set $\Sigma$, is $\Sigma$ satisfiable?
    \item \emph{Implication} - Given a set of GGDs $\Sigma$ and a GGD $\sigma$, does $\Sigma$ imply $\sigma$, i.e. is $\sigma$ a logical consequence of $\Sigma$, (denoted by $\Sigma \models \sigma$) for every non-empty graph G that satisfies $\Sigma$? 
\end{itemize}

\subsection{Validation}
\label{sec:validation}

We first study the \textbf{validation problem} for GGDs. This problem has already been discussed in~\cite{Shimomura2020}. 
The \textbf{validation problem} for GGDs is defined as: Given a finite set $\Sigma$ of GGDs and graph G, does $G \models \Sigma$ (i.e., $G \models \sigma$ for each $\sigma\in\Sigma$)? 
An algorithm to validate a GGD $\sigma = Q_s[\overline{x}], \phi_s \rightarrow Q_t[\overline{x},\overline{y}],\phi_t$ was proposed in~\cite{Shimomura2020}. This algorithm returns true if the $\sigma$ is validated and returns false if $\sigma$ is violated.  
Algorithm~\ref{algo:validation} summarizes the steps of the validation algorithm. 

\begin{algorithm}[htb]
    \caption{Validation of Graph Generating Dependency}
    \label{algo:validation}
    \begin{algorithmic}[1] 
        \Procedure{Validation}{$\sigma = Q_s[\overline{x}]\phi_s \rightarrow Q_t[\overline{x},\overline{y}]\phi_t$, Graph $G$}
            \State Search for matches of the source graph pattern, $\llbracket Q_s[\overline{x}] \rrbracket_G$  
            \For{ each match $h_s[\overline{x}] \in \llbracket Q_s[\overline{x}]\rrbracket_G$} 
            \If {$h_s[\overline{x}]$ satisfies the source differential constraints (ie., $h_s[\overline{x}] \models \phi_s$)}
            \State Search for matches of the target graph pattern $\llbracket Q_t[\overline{x},\overline{y}] \rrbracket_G$ where for a match $h_t(\overline{x},\overline{y})$ for all $x\in\overline{x}$ there is $h_s(x) = h_t(x)$. 
            \If{$\llbracket Q_t[\overline{x},\overline{y}] \rrbracket_G = \emptyset$}
            \State \textbf{return} false
            \EndIf
            \If {$\not\exists h_t[\overline{x},\overline{y}] \in \llbracket Q_t[\overline{x},\overline{y}] \rrbracket_G \models \phi_t$}
            \State \textbf{return} false
            \EndIf
            \EndIf
            \EndFor
        \State \textbf{return} true \Comment{If the for loop ended without returning a value then it means that the GGD is validated}    
        \EndProcedure
    \end{algorithmic}
\end{algorithm}

Algorithm~\ref{algo:validation} is repeated for each $\sigma \in \Sigma$. 
For each match on which $\sigma$ is violated, new vertices/edges can be generated in order to repair it (i.e, in order to make the GGD $\sigma$ valid on $G$).

A problem in $\Pi_{2}^P$ complexity is a problem solvable by a nondeterministic Turing machine with an oracle for a coNP problem~\cite{papadimitriou1994computational}. 
Given the complexity of evaluation of the source and target side of a GGD and its semantics, to prove the complexity of the validation problem for GGDs we can directly reduce it to a $\forall$-QSAT problem by adapting the reduction presented in~\cite{Pichler2011} for the evaluation of \tgds.

Graph pattern matching queries can be expressed as conjunctive queries (CQ)~\cite{Bonifati2018} which are well-known to have NP-complete evaluation complexity~\cite{Pichler2011}. This complexity has been proven in the literature by reducing the graph pattern matching problem into a SAT problem~\cite{Hell1990,Kirousis2001,Feder1998}.
Considering that the complexity of evaluating the set of source and target differential constraints, $\phi_s$ and $\phi_t$, is in P, the the cost of evaluating if a graph pattern match satisfies a set of differential constraints is linear to the number of differential constraints in this set. 
Hence, overall, the complexity of evaluating the source and target constraints of a GGD separately is dominated by the complexity of the evaluation of the graph patterns.

The $\forall$-QSAT problem has been proven to be in $\Pi_{2}^P$ complexity~\cite{Pichler2011,papadimitriou1994computational,arora2009computational}, hence, the following theorem. 

\begin{theorem}
The validation algorithm of GGDs is in $\Pi_{2}^P$ complexity.
\end{theorem}

The graph pattern matching problem is solvable in PTIME when the graph pattern has bounded treewidth~\cite{Fan2016,Pichler2011}, which corresponds to graph patterns covering over 99\% of graph patterns observed in practice~\cite{BonifatiMT20}. 
Suppose that both source and target graph pattern matching are both treewidth bounded. In this case, we can substitute the NP complexity of evaluating the graph patterns by a P complexity, and the overall complexity in this case drops to coNP complete (see similar proof in~\cite{Pichler2011}). 

\subsection{Satisfiability}
\label{sec:satisfiability}

A set of GGDs $\Sigma$ is satisfiable if: (i) there exists a graph $G$ which is a model of $\Sigma$ ($G \models \Sigma$) and (ii) for each GGD $\sigma \in \Sigma$ there exists a match of $Q_s[\overline{x}]$ in $G$.
Then, the satisfiability problem is to determine if a given set of GGDs $\Sigma$ is satisfiable.

Before considering the general satisfiability problem as outlined above, we discuss if a single GGD $\sigma$ of the set $\Sigma$ is satisfiable or not. The satisfiability of a GGD $\sigma$ depends on the satisfiability of the sets of differential constraints $\phi_s$ and $\phi_t$ and not on the topological constraints (source and target graph patterns). 

The satisfiability problem for differential constraints was studied by~\cite{Song2011} in the context of differential dependencies (DDs) for relational databases.
A set $\phi$ of differential constraints is unsatisfiable if the intersection of any two differential constraints in $\phi$ is infeasible (see infeasibility definition in Section~\ref{sec:properties}). Observe an example of this case in Example~\ref{example:sat1}.

Assuming that the set of source (and target) differential constraints $\phi_s$ (and $\phi_t$) of each GGD $\sigma$ in the set $\Sigma$ is satisfiable,  we discuss the satisfiability of the \emph{set} of GGDs $\Sigma$, as defined at the beginning of the section. Informally, the satisfiability problem is to check if the GGDs defined in $\Sigma$ conflict with each other. Similar to the previously proposed graph dependencies~\cite{Fan2016,Fan2019a}, a set $\Sigma$ of GGDs may \textbf{not} be satisfiable. 

    \begin{figure}[t]
  \centering
  \includegraphics[width=0.8\linewidth]{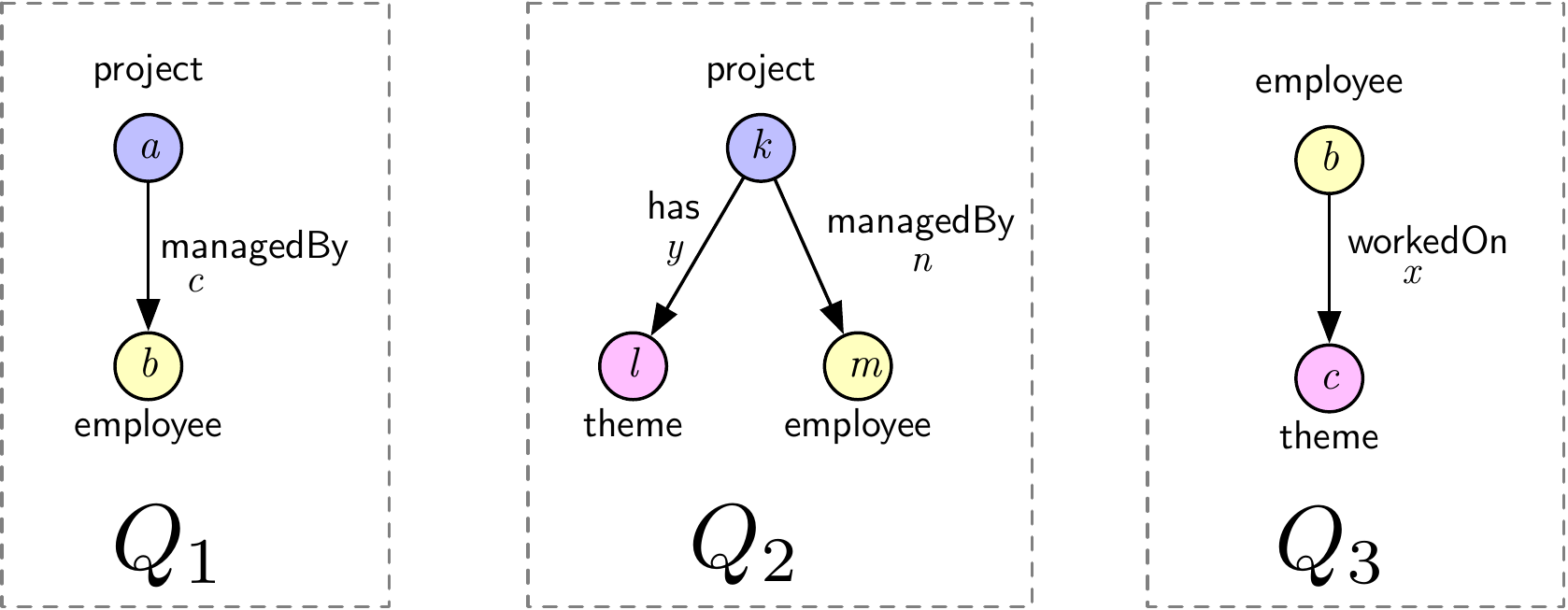}
  \caption{Graph patterns in GGDs.}
  \label{fig:exampleSatisfiability}
\end{figure}

\begin{example}\label{example:sat1}
Consider the graph pattern $Q_1$ shown in \autoref{fig:exampleSatisfiability}. 
Suppose $\phi_s=\{(\delta_{type}(b.type, ``full-time") = 0)$, $(\delta_{type}(b.type, ``full-time") > 1)\}$. 
Observe that the differential constraints in $\phi_s$  are disjoint.
This means that it is not possible to agree on all differential constraints of $\phi_s$, so $\phi_s$ is infeasible.
\end{example}

\begin{example}\label{exampleSat2}
Consider the graph patterns shown in \autoref{fig:exampleSatisfiability} and the following two GGDs: $\sigma_1 = Q_1[a,b,c]\{\emptyset\} \rightarrow Q_3[b,c,x]\{\delta_A(b.A, \alpha)>10\}$ and $\sigma_2 = Q_2[k,l,m,n,y]\{\emptyset\} \rightarrow Q_3[b,c,x]\{\delta_A(b.A, \alpha) \le 10\}$ in which $\alpha$ is a constant value. There might exist matches of $Q_1$ and $Q_2$ that map to the same nodes and edges for the labels \{project, managedBy, employee\}, even though this implies the same target graph pattern, the constraints $\phi_t$ of $\sigma_1$ and $\sigma_2$ are disjoint. In consequence, these two GGDs are not satisfiable. 
\end{example}

GGDs defined with different graph patterns may interact with each other. Additional challenges of the GGDs satisfiability compared to the proposed graph dependencies in the literature are: (i) dealing with additional complexity of GGDs target graph pattern $Q_t[\overline{x},\overline{y}]$ which is an extension of the source graph pattern and (ii) the checking of the feasibility of differential constraints in $\phi_t$ in the cases in which the graph pattern variables are mapped to the same nodes or edges of the graph $G$. To analyse the satisfiability of a set of GGDs, we use the Chase procedure presented in Section~\ref{sec:chase}.

A set $\Sigma$ is said to be satisfiable if there exists a graph $G$ which is a model of $\Sigma$ ($G \models \Sigma$) and for each GGD $\sigma \in \Sigma$ there exists a match of $Q_s[\overline{x}]\phi_s$ in $G$.
To verify the satisfiability of $\Sigma$, we build a canonical graph $G_\Sigma$ which contains one match of the source side of each one of the GGDs in $\Sigma$. The graph $G_\Sigma$ is defined as $(V_\Sigma,E_\Sigma,\lambda_\Sigma,\eta_\Sigma, \nu_\Sigma)$. Each vertex $v \in V_\Sigma$ and each $e \in E_\Sigma$ are a tuple of $(l, v, [v.R_q])$ in which $l$ and $v$ is the label and the variable alias of the vertex/edge, respectively, and $[v.R_q]$ is the set of range classes attached to this vertex/edge ($[v]_{R_q}$ and attribute range classes, i.e. $[v.A]_{R_q}$).
For each variable $x$ in $Q_s[\overline{x}]\phi_s $ of $\sigma \in \Sigma$, we add $x$ to $G_\Sigma$ according to the following steps:
\begin{enumerate}
    \item If $G_\Sigma$ is empty: (a) For each variable $x \in \overline{x}$, create a range class for the variable $x$, $[x]_{R_q}$. Search for differential constraints in $\phi_s$ that involve the variable $x$. For each of one of the differential constraints, create a $rcq$ and add it to $[x]_{R_q}$. In case the $rcq$ refers to a property of $x$, i.e. $x.A$, create and add the $rcq$ to $[x.A]_{R_q}$. 
    (b) If $x \in V_Q$ of $Q_s[\overline{x}]$, add a node $x' = (\lambda_Q(x), x, [x.R_q]$) to $V_\Sigma$, $V_\Sigma = V_\Sigma \cup x'$.
    (c) If $x \in E_Q$ of $Q_s[\overline{x}]$, add an edge $x' = (\lambda_Q(x), x, [R_q]$) to $E_\Sigma$, $E_\Sigma = E_\Sigma \cup x'$.
    \item If $G_\Sigma$ is not empty: (a) Check if $Q_s[\overline{x}]\phi_s$ interacts with $G_\Sigma$. (b) If there are no interactions, then repeat step 1 for $Q_s[\overline{x}]\phi_s$. 
    (c) If there are vertices and edges in $G_\Sigma$ that interact which $Q_s[\overline{x}]\phi_s$, for each variable $x \in Q_s[\overline{x}]\phi_s$ that interacts with $G_\Sigma$ to a node or edge which the variable alias is $x'$, rename the variable $x$ to $x'$ in $\phi_s$ then update the set $[x'.R_q]$ in $G_\Sigma$. 
\end{enumerate}

\begin{theorem}\label{th:sat}
A set $\Sigma$ is satisfiable if and only if every Chase step of the Chase procedure over $G_\Sigma$ is consistent and the Chase terminates.
\end{theorem}

\paragraph{Proof Sketch} - First, we show that given a set of consistent Chase steps, there exists a graph $G$ that can be derived from $G_\Sigma$ and is a model of $\Sigma$.

We assume that the Chase terminates and prove that every Chase step should be consistent for a set to be Satisfiable. By contradiction, assume that there exists a Chase step $(\sigma_i, h_{s_i}[\overline{x}])$ on $G_\Sigma$, in which $\sigma_i \in \Sigma$ and $h_{s_i}[\overline{x}]$ is a match of the source on $G_\Sigma$, that is not consistent, we try to prove that $\Sigma$ is satisfiable. 
If the Chase step $(\sigma_i, h_{s_i}[\overline{x}])$ is not consistent, according to the Chase procedure it means that (1) there is a label conflict or (2) the enforced differential constraints $\phi_t$ of $\sigma_i$ are not feasible with the range classes of the matched nodes or edges in $G_{\Sigma}$. If there is a label conflict then there exists two nodes/edges $x$ and $x'$ in $G_\Sigma$, $\lambda_\Sigma(x) \neq \lambda_\Sigma(x')$ and consequently $x \neq x'$, therefore the set $\Sigma$ is not satisfiable, contradicting our assumption.
If the enforced differential constraints are not feasible then there are no values that can satisfy all the enforced constraints therefore the set $\Sigma$ is not satisfiable (see consistency on differential constraints~\cite{Song2011}), contradicting our assumption.

If $\Sigma$ is satisfiable and Chase terminates and every Chase step is consistent, assuming that $G_\Sigma'$ is the final graph after applying the Chase steps, we show that there exists a graph $G$ that can be derived from $G_\Sigma'$ that is a model of $\Sigma$.
For each $u \in (V \cup E)$ in $G_\Sigma'$:
\begin{enumerate}
    \item if $u \in V$, create a node $v$ with the same label as $u$ and add it to $G$. Respectively, if $u \in E$, create an edge $e$ with the same label and connect the same nodes as $u$ in $G_\Sigma$ and add it to $G$. Next, we populate $v$ or $e$ attributes;   
    \item for each node or edge $x$ in $[u]_{R_q}$, update $x$ attribute range classes with $[u.A]_{R_q}$. This means for every attribute $A$ in $x$ and $u$, $[u.A]_{R_q} = [x.A]_{R_q}$;
    \item for each attribute $A$ range class $[u.A]_{R_q}$, assign a distinct constant values $c$ to $u.A$ such that $c \models [u.A]_{R_q}$. That means, assign a value $c$ that can satisfy all the enforced differential constraints that are represented in $[u.A]_{R_q}$.  
\end{enumerate}

Next, it suffices to show that the graph $G$ derived from $G_\Sigma'$, $G \models \Sigma$. 
Assume by contradiction that there exists a GGD $\sigma = Q_s[\overline{x}]\phi_s \rightarrow Q_t[\overline{x},\overline{y}]\phi_t \in \Sigma$ such that $\sigma \not\models G$, that is, exists a match $h_s[\overline{x}] \models \phi_s$ of $\sigma$ in $G$ but there does not exist a match of the target $h_t[\overline{x},\overline{y}]\phi_t$ of $\sigma$ in $G$. 
Since $G_\Sigma$ is built according to the source graph pattern of $\Sigma$, we can ensure that all GGDs in $\Sigma$ will be applied at least once in $G_\Sigma$, if every Chase step is consistent and terminates, it means that $G_\Sigma' \models \Sigma$. Given the $G$ construction procedure, we can conclude that $G$ and $G_\Sigma'$ are isomorphic and consequently $G \models \Sigma$, contradicting our assumption that there exists a GGD $\sigma$ such that $\sigma \not\models G$.

Now, assume by contradiction that the Chase does not terminate and the set $\Sigma$ is satisfiable. 
We are also assuming that every Chase step is consistent, as if a Chase step is inconsistent then the Chase procedure terminates.
Given this assumption, if the Chase does not terminate, then, according to the procedure in Section~\ref{sec:chase}, it means that there is always a GGD $\sigma \in \Sigma$ than can be applied and the Chase step will either update or generate new nodes or edges in $G_\Sigma$. In practice, it means that there is always a match of the source of $\sigma$ in $G_\Sigma$ that is not validated. As a consequence, given that it is possible to derive a model graph $G$ from $G_\Sigma$, if we construct $G$ from $G_\Sigma$ at any point of the non-terminating Chase procedure the built $G \not\models \Sigma$ and therefore $\Sigma$ is not satisfiable, contradicting our assumption.

Assume by contradiction that $\Sigma$ is not satisfiable, the Chase terminates and every Chase step on $G_\Sigma$ is consistent. Since $\Sigma$ is not satisfiable it means that there does not exist a graph model $G$ such that $G \models \Sigma$. If the model graph does not exist it also means that either does not exist $G_\Sigma$ or it is not possible to derive a graph $G$ from the canonical graph $G_\Sigma$. Considering that the Chase terminates, if it is not possible to derive a graph $G$ from $G_\Sigma$, it is not possible to satisfy all enforced differential constraints for all nodes or edges to be created in $G$ from $G_\Sigma$, which means that there are range classes in nodes or edges that are not feasible. From this, we can affirm that there exists a Chase step that is not consistent, contradicting our assumption.
If $G_\Sigma$ does not exist, then it means that the set $\Sigma$ is empty, as $G_\Sigma$ is initially built from the source graph patterns of the GGDs in $\Sigma$. If the set $\Sigma$ is empty, then it is not possible to apply the Chase procedure therefore there does not exists any Chase step, contradicting our assumption.

\subsection{The Satisfiability Problem for Weakly-Acyclic GGDs}

As mentioned previously, the Chase for GGDs might not always terminate (see Section~\ref{para:chaseTermination}).
Given the relationship between \tgds and GGDs, we can also affirm that deciding if the Chase will terminate for a set of GGDs is also undecidable. As presented, knowing if the Chase can terminate is key for checking the satisfiability of a set of GGDs and therefore we can deduce that the satisfiability problem for GGDs is also undecidable. 

However, research on \tgds has identified that if a set of \tgds is weakly-acyclic (see Subsection~\ref{para:chaseTermination}) then all Chase versions and all sequences of \tgds will terminate~\cite{Fagin2005, greco2012}. We can adapt such results from \tgds on GGDs, as a consequence, if a set of GGDs is weakly-acyclic then the Chase procedure proposed terminates.

\begin{theorem}\label{th:satisfiabilitycomplexity}
The satisfiability problem of weakly-acyclic GGDs is in coNP.
\end{theorem}

Before checking the satisfiability, we need to check if a set of GGDs is weakly-acyclic to ensure the termination of the Chase procedure. 
To check if a set of GGDs is weakly acyclic we adapt the proposed algorithm in PTIME by the authors in~\cite{Fagin2005} for checking if a set of \tgds is weakly acyclic. The main difference is that GGDs also have differential constraints, so for building the dependency graph proposed in~\cite{Fagin2005}, each node refers to an entity, and the set of differential constraints referring to that entity according to a set of GGDs. The alterations needed on the dependency graph construction do not increase the complexity of the algorithm, therefore, the complexity remains in PTIME.  

\begin{lemma}\label{le:ggdstermination}
The complexity of checking if a set of GGDs is weakly-acyclic is in PTIME.
\end{lemma}

\paragraph{Proof Sketch of \autoref{th:satisfiabilitycomplexity}} - 
We can summarize the process described for checking satisfiability of a set $\Sigma$ as an coNP algorithm (an algorithm that checks disqualifications in polynomial time) in which the two main steps are: (1) Build a canonical graph $G_\Sigma$ and, (2) given any GGD $\sigma \in \Sigma$, apply each existing match $h_s[\overline{x}]$ of the source graph pattern as a Chase step $chase(h_s[\overline{x}],\sigma)$ in $G_\Sigma$ and check if the Chase step is consistent or not. If there exists a Chase step that is not consistent, the algorithm rejects $\Sigma$ (according to \autoref{th:sat}).

To prove its complexity is in coNP, we analyse each step of the algorithm.
To build the canonical graph $G_\Sigma$, we need to check the interaction on the source of the GGDs in $\Sigma$ (see interaction definition on Section~\ref{sec:properties}). The interaction on the source (or the target) of the two GGDs can be defined as the maximal subgraph pattern in which the source graph patterns of these GGDs have in common. The maximal subgraph problem is a problem studied and proven to be NP-hard~\cite{Lewis78}.

Assuming that a GGD $\sigma$ can be applied to $G_\Sigma$ (there exists a match of the source graph pattern of $\sigma$ in $G_\Sigma$), at each Chase step we enforce the target side of $\sigma$ either by generating new nodes and edges and initializing its range classes (in case a match of the target does not exist) or by updating the existing range classes of the matched nodes and edges according to $\phi_t$ of $\sigma$. 

A Chase step is consistent if for all existing matches of the source of $\sigma$ and matches of the target of $\sigma$ (generated or not), the range classes of the matched nodes and edges are feasible.
Given a GGD $\sigma$ and $G_\Sigma$, we can rewrite the problem of checking if Chase steps are consistent as:
\begin{equation}\label{qbfChase}
    \forall (h_s[\overline{x}] \wedge h_t[\overline{x},\overline{y}] (R_q(h_t[\overline{x},\overline{y}])))
\end{equation}

in which, abusing of the notation, $R_q(h_t[\overline{x},\overline{y}])$ refers to the range classes of the matched nodes and edges of $h_t[\overline{x},\overline{y}]$.
It is known that matching a graph pattern to a graph $G$ is equivalent to evaluating a conjunctive query in $G$, which is proven to be in NP~\cite{Chandra1977} by a reduction to a 3-SAT problem.
Checking if range classes are infeasible is comparable to checking if two differential constraints are feasible. According to~\cite{Song2011}, the complexity of checking if two differential functions on the same attribute of a relation are feasible is in P by a reduction to a 2-SAT problem. 
Given these results, we can reduce (\ref{qbfChase}) to a TAUTOLOGY problem. 
A TAUTOLOGY problem is defined as, given a Boolean formula, determine if every possible value assignment to variables of this formula results in a true statement~\cite{arora2009computational}.
The TAUTOLOGY problem is proven to be in coNP, proving that the complexity of the satisfiability problem is in coNP.

\subsection{Implication}
\label{sec:implication}

The \textbf{implication problem} for GGDs is defined as: assuming a finite set $\Sigma$ of GGDs and a GGD $\sigma =  Q_{s}[\overline{x}], \phi_{s} \rightarrow Q_{t}[\overline{x},\overline{y}]\phi_{t}$, does $\Sigma \models \sigma$?

The implication problem is to define whether the GGD $\sigma$ is a logical consequence of the set $\Sigma$. 
We use implication to minimize dependencies and optimize data quality rules~\cite{Fan2019a}.
Here, we assume that the GGD set $\Sigma$ and the GGD $\sigma$ are \textit{satisfiable}, and the set $\Sigma$ is irreducible. That means that there is no GGD $\sigma_\alpha \in \Sigma$ that can be implied from the set $\{\Sigma - \sigma_\alpha\}$. Note that since we are assuming that $\{\Sigma \cup \sigma\}$ are satisfiable, the Chase procedure for this set of GGDs always terminates (see Section~\ref{sec:satisfiability}).

\begin{figure}[t]
  \centering
    \includegraphics[width=0.95\linewidth]{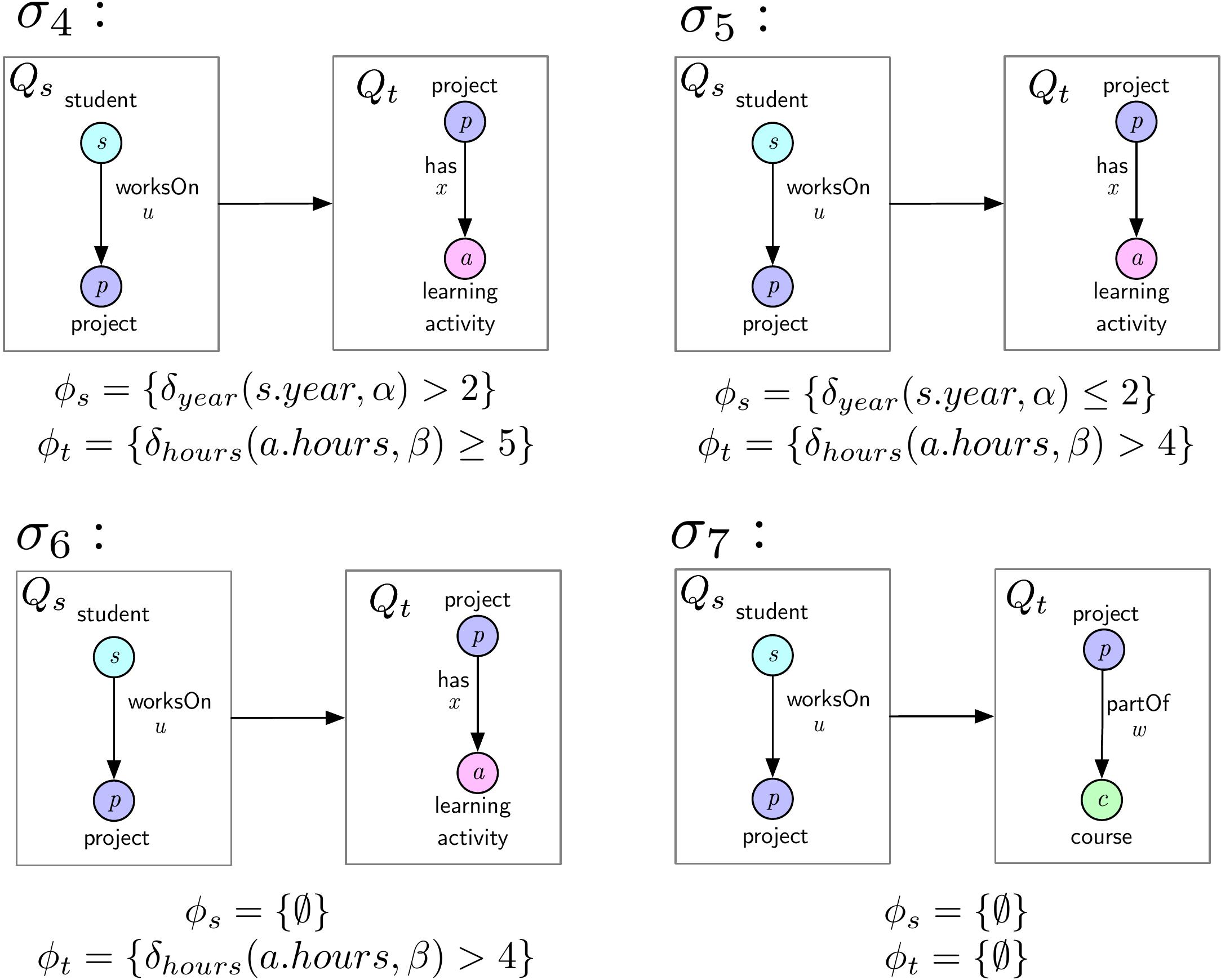}
  \caption{Implication Example in GGDs}
  \label{fig:exampleImplication}
\end{figure}

Similar to the satisfiability problem, the implication needs to consider the interaction between the GGD $\sigma$ and the GGDs in $\Sigma$. 
However, the satisfiability problem only needs to verify if there are any conflicts regarding the source variables on the target side of the GGDs. On the other hand, to solve the implication problem, we need to verify if it is possible to derive/imply $\sigma$ source and target constraints from the GGDs in $\Sigma$. 

To solve the implication problem, we use the Chase procedure we proposed earlier. Here, we build a graph named $G_{closure}$ and use it as a tool to identify if the implication holds. In $G_{closure}$, similar to $G_\Sigma$ in Section~\ref{sec:satisfiability}, each node and edge of $G_{closure}$ is a triple $(l,v,[v.R_q])$. 
We initialize $G_{closure}$ with the source graph pattern of $\sigma$ using the same procedure proposed to build $G_\Sigma$.
The main idea here is to use the Chase method to enforce the GGDs of $\Sigma$ on $G_{closure}$ and check if all possible matches of the target side of $\sigma$ are covered in $G_{closure}$. If this is true, then it means that the same result of $\sigma$ is enforced by the GGDs in $\Sigma$.

We say that $\sigma$ is deducible from $G_{closure}$, if there exists a match of $Q_t[\overline{x},\overline{y}]\phi_t$ of $\sigma$ in $G_{closure}$ in which the range classes of the matched nodes/edges subsumes the differential constraints in $\phi_t$.
Observe that in $G_{closure}$ we do not set attribute values but instead use range classes that express all the possible values an attribute can be assigned to. 

\begin{theorem}
$\Sigma \models \sigma$ if and only if there exists a sequence of consistent Chase steps on Chase($G_{closure}, \Sigma$), and $Q_t[\overline{x},\overline{y}]\phi_t$ of $\sigma$ is deducible from $G_{closure}$.
\end{theorem}

\paragraph{Proof Sketch}
Assuming that $\Sigma \models \sigma$, to prove the above theorem we first consider two cases and prove it by contradiction: (1) there does not exist a sequence of consistent Chase steps and (2) there exists a sequence of consistent Chase steps and $Q_t[\overline{x},\overline{y}]\phi_t$ is not deducible from $G_{closure}$.
If there does not exist a sequence of consistent Chase steps and $\{\sigma, \Sigma\}$ is satisfiable (there is a sequence of consistent Chase steps in $G_{\Sigma}$, see Section~\ref{sec:satisfiability}) then it means that there are no $\sigma_i \in \Sigma,$ in which there a match of the source of $\sigma_i, h_{s_i}[\overline{x}]$ that can be applied as a Chase step in $G_{closure}$. Therefore, $\forall \sigma_i \in \Sigma, \not\exists h_{s_i}[\overline{x}]$ in $G_{closure}$ and $\Sigma \not\models \sigma$, contradicting our first assumption.

Considering that there exists a sequence of consistent Chase steps, we prove that if we can deduce the target of $\sigma$ from $G_{closure}$, that means for any possible graph $G$ in which the set $\Sigma$ is valid, the set of matches of the target of $\sigma$ are contained in $G_{closure}$.
We prove it by contradiction, so assuming that $\Sigma \models \sigma$ and there exists a sequence of consistent Chase steps $((\sigma_0, h_{s_0}[\overline{x}]), ..., (\sigma_x, h_{s_x}[\overline{x}]))$ on $G_{closure})$ in which $(\sigma_0, ...., \sigma_x) \in \Sigma$, we prove that there exists a graph $G$ that $(\sigma_0, ...., \sigma_x) \models G$ and  $\llbracket Q_t[\overline{x},\overline{y}]\phi_t \rrbracket_G \not\sqsubseteq G_{closure}$. 

If $\Sigma \models \sigma$, and there exists a sequence of consistent Chase steps, then $G_{closure} \models \{\Sigma \cup \sigma\}$ (see Section~\ref{sec:satisfiability}). That means $G_{closure}$ contains at least one match of the source side of $\sigma$ and there exists a match $h_{t_\sigma}[\overline{x},\overline{y}]$ of the target side of $\sigma$ in $G_{closure}$. Since $\Sigma \models \sigma$, then according to the Chase procedure, the range classes of the nodes and edges in the match $h_{t_{\sigma}}[\overline{x},\overline{y}]$ subsume $\phi_t$. This subsumption means that every possible value in $\phi_t$ it is covered by the range classes of the match $h_{t_{\sigma}}[\overline{x},\overline{y}]$ in $G_{closure}$. Conclusively, for every graph $G$ that $\Sigma \models G$ any matches of the target side of $\sigma$ in $G$ can be represented by the match $h_{t_{\sigma}}[\overline{x},\overline{y}]$ in $G_{closure}$. 

Assume that there exists a sequence of consistent Chase steps and the target $Q_t[\overline{x},\overline{y}]\phi_t$ is deducible from $G_{closure}$ and $\Sigma \not\models \sigma$, we prove by contradiction that this is not correct. Given the Chase procedure and that there exists a sequence of consistent Chase steps, we can affirm that GGDs from the set $\Sigma$ were applied to $G_{closure}$ and $\sigma$ was not. Now, if the target of $\sigma$ is deducible from $G_{closure}$ it means that for every match of the source of $\sigma$ in $G_{closure}$ there exists a match of the target of $\sigma$ in $G_{closure}$, if $\sigma$ was not applied during Chase we can confirm that $\Sigma \models \sigma$, contradicting our assumption that $\Sigma \not\models \sigma$.

\begin{figure*}[t]
  \centering
    \includegraphics[width=0.8\linewidth]{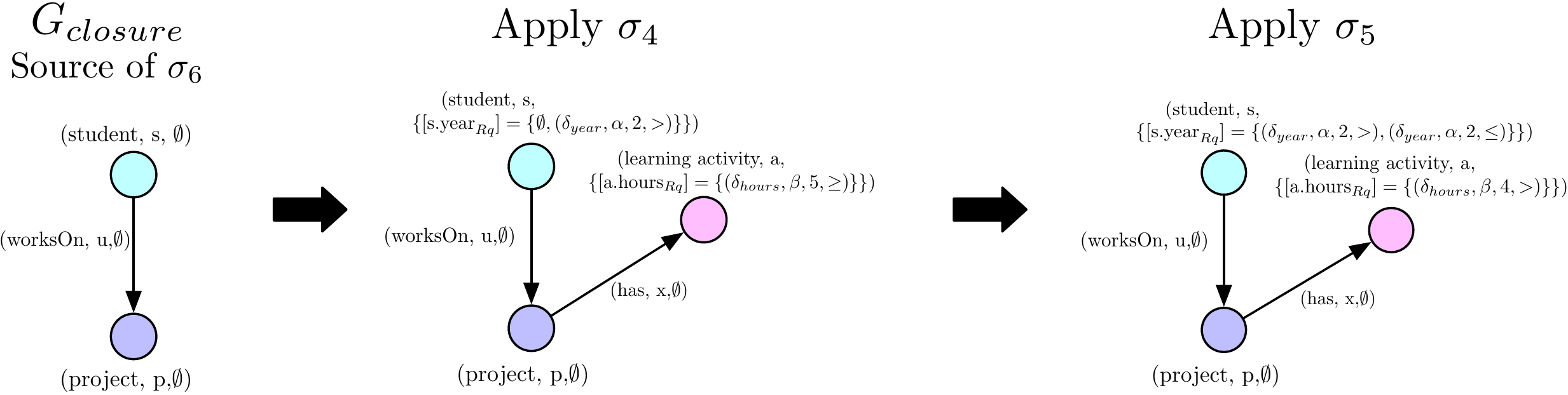}
  \caption{Chase procedure on \autoref{fig:exampleImplication}}
  \label{fig:implicationChase}
\end{figure*}

\begin{example}
Given the GGDs in \autoref{fig:exampleImplication} in which $\alpha$ and $\beta$ are two constant values, does $\Sigma = \{\sigma_4,\sigma_5\} \models \sigma_6$? Observe \autoref{fig:implicationChase} in which we apply the Chase procedure to solve this problem. For the means of presentation, in the figure, we show only the range classes of properties used during Chase.

To verify if $\{\sigma_4,\sigma_5\} \models \sigma_6$, the first step is to initialize $G_{closure}$ with the source side of $\sigma_6$, since the $\phi_s = \{\emptyset\}$ in $\sigma_6$, we initialize all range classes with $\emptyset$. 
The next step is to apply a GGD in $\{\sigma_4, \sigma_5\}$ to $G_{closure}$. In this case, both GGDs $\sigma_4$ and $\sigma_5$ are applicable, so we can choose to apply $\sigma_4$ first. Given that the source constraint $\phi_s = \{\delta_{year}(s.year, \alpha) > 2\}$ is subsumed by the range class of student in $G_{closure}$, we create a $rcq$ for this constraint and add it to the range class of student. Next, we check if there exists a match of the target of $\sigma_4$. Since the target does not exist, we create a new edge and a new node and initialize its range classes according to $\phi_t$ of $\sigma_4$. 

Next, we use the same process and apply $\sigma_5$, observe that the range class of $s.year$ subsumes the source constraint of $\sigma_5$ because of the $\emptyset$, making $\sigma_5$ possible to apply. Following the same procedure as in $\sigma_4$, we create a $rcq$ for $\phi_s$ and add it to the $s.year$ range class. Next, we check the target of $\sigma_5$ in $G_{closure}$, which we can easily verify that there exists a match and that the range class of the node labeled "learning activity" is subsumed by the $\phi_t$ in $\sigma_5$. Given the subsumption property, we update the range class of the node learning activity with the constraint in $\phi_t$. In this case, the Chase terminates as applying any GGDs in $\{\sigma_4, \sigma_5\}$ will not cause any updates on $G_{closure}$. Given the final $G_{closure}$, we can easily verify that the answer to this implication is true as it is possible to deduce $\sigma_6$ from $G_{closure}$. 
\end{example}

\begin{figure*}[t]
  \centering
  \includegraphics[width=0.6\linewidth]{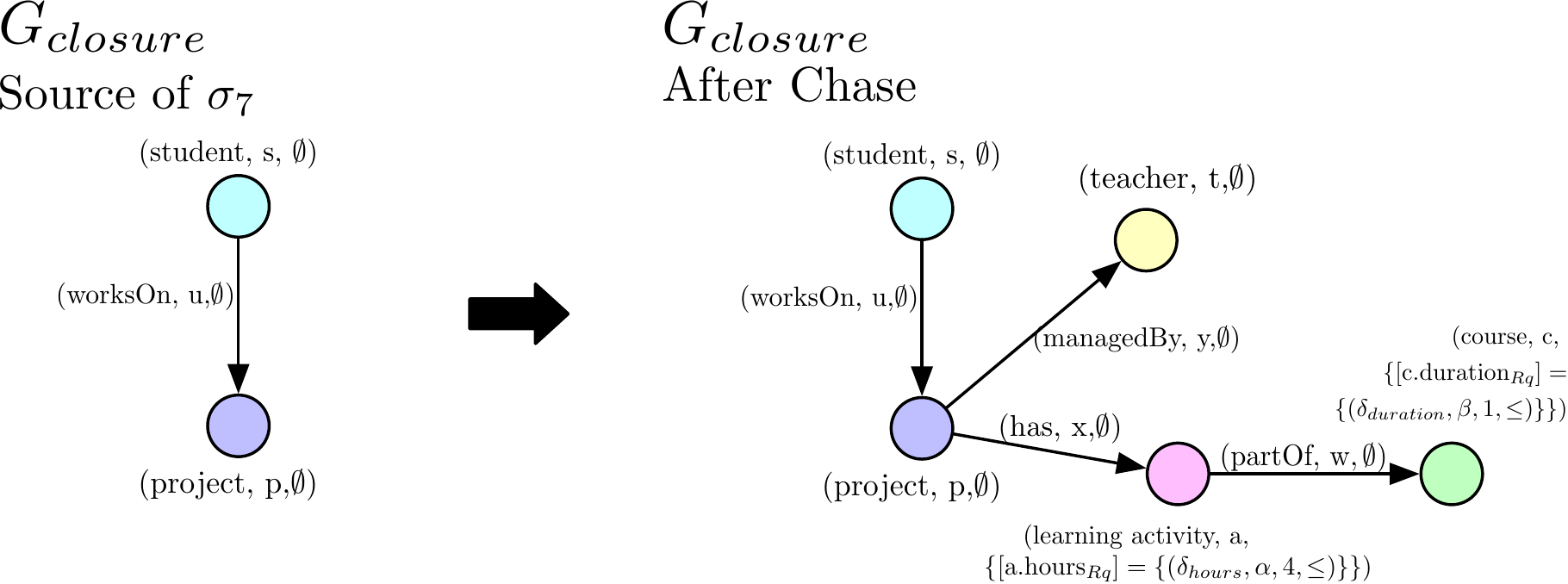}
  \caption{Chase on Transitive GGDs}
  \label{fig:transitiveChase2}
\end{figure*}

\begin{example}
In \autoref{fig:exampleImplication}, we used as an example a set of GGDs with the same source and target graph patterns. Here, we show an example of when we do not have the same graph patterns. Given the GGDs $\sigma_2$ and $\sigma_3$ in \autoref{fig:transGGDs}, we rewrite it with differential constraints $\sigma_{2'} = \phi_t=\{\delta_{hours}(a.hours,\alpha) \le 4\}$ and $\sigma_{3'} = \phi_s = \{\delta_{hours}(a.hours,\alpha) \le 8\}, \phi_t = \{\delta_{duration}(c.duration,\beta) \le 1\}$, does $\{\sigma_{2'},\sigma_{3'}\} \models \sigma_{7}(\phi_s = \{ \emptyset \}, \phi_t = \{\delta_{duration}(c.duration,\beta) \le 2\})$ (\autoref{fig:exampleImplication})? 
Observe in \autoref{fig:transitiveChase2} the initial $G_{closure}$ and the final $G_{closure}$ after the Chase procedure. From the final version of $G_{closure}$ we can verify that $\{\sigma_{2'}, \sigma_{3'}\} \models \sigma_7$ is true.
\end{example}

\begin{theorem}
The implication problem for GGDs is in coNP.
\end{theorem}
 
\paragraph{Proof Sketch} - We can summarize the steps of the complexity algorithm as:

\begin{enumerate}
    \item Build $G_{closure}$.
    \item Execute the Chase procedure $Chase(\Sigma, G_{closure})$ until it terminates and verify if each Chase step is consistent. If a Chase step is not consistent, then return false and reject $\Sigma$.
    \item Verify if it is possible to deduce the target side of $\sigma$ from the resulting $G_{closure}$. If yes, then return true. Otherwise, return false and reject $\Sigma$.
\end{enumerate}

To analyse the complexity of this Chase procedure, we analyse the complexity of the each step.
$G_{closure}$ is initialized by just the source graph pattern of $\sigma$, which means that there is no need to check for interaction in case $G_{closure}$ is not empty and consequently the initialization of $G_{closure}$ is in PTIME. 
After initializing $G_{closure}$, the next step is to apply $Chase(\Sigma, G_{closure})$ and check if each Chase step is consistent or not. This step are the exact same steps as shown in the proof of \autoref{th:satisfiabilitycomplexity} and proved to be in coNP. 

If there are only consistent steps, it suffices to check if the target of $\sigma$ can be deduced from the resulting $G_{closure}$ after Chase terminates (observe that we assume that $\{\Sigma \cup \sigma\}$ is satisfiable and, therefore, the Chase procedure terminates for this set). This procedure is comparable to checking if there exists a match of the target of $\sigma$ in $G_{closure}$ which can be done in polynomial time. If it does not exist, then the implication is false and $\Sigma$ is rejected. Therefore, the implication problem is in coNP.

\section{GGDs for Data Inconsistencies: An Experimental Study}

In this section, we discuss and demonstrate the practical use and the feasibility of GGDs to find data inconsistencies (matches of the source graph pattern that are not validated). Finding inconsistent data is the first step to identifying data in the graph that should be repaired. 

Given a set of GGDs $\Sigma$, we define as inconsistent data according to $\Sigma$ as: set of graph pattern matches of the source side of each GGD in which the target does not hold.

In the following subsections, we give details on the implementation and experiments setup (Subsection~\ref{sec:experiments}) and an analysis on the impact in execution time according to different aspects of using GGDs.

\subsection{Implementation and Experiment Set Up}
\label{sec:experiments}

From the definition of inconsistent data we can observe that this problem is related to check if a set of GGDs, $\Sigma$, is violated or not. For this reason, to identify inconsistent data, we modify the previously introduced validation algorithm (Algorithm~\ref{algo:validation}) to, instead of returning true or false if the set is validated or not, return which matches of the source ($h_s(Q_s[\overline{x}) \models \phi_s$) were not validated. Algorithm~\ref{algo:dataInconsistency} presents the modified validation algorithm for identifying data inconsistencies.

\begin{algorithm}[htb]
    \caption{Inconsistent Data Identification according to GGDs}
    \label{algo:dataInconsistency}
    \begin{algorithmic}[1] 
        \Procedure{DataInconsistency}{$\sigma = Q_s[\overline{x}]\phi_s \rightarrow Q_t[\overline{x},\overline{y}]\phi_t$, Graph $G$}
            \State Violated $\gets \emptyset$ \Comment{Set of Violated source matches}
            \State Search for matches of the source graph pattern, $\llbracket Q_s[\overline{x}] \rrbracket_G$  
            \For{ each match $h_s[\overline{x}] \in \llbracket Q_s[\overline{x}]\rrbracket_G$} 
            \If {$h_s[\overline{x}]$ satisfies the source differential constraints (ie., $h_s[\overline{x}] \models \phi_s$)}
            \State Search for matches of the target graph pattern $\llbracket Q_t[\overline{x},\overline{y}] \rrbracket_G$ where for a match $h_t(\overline{x},\overline{y})$ for all $x\in\overline{x}$ there is $h_s(x) = h_t(x)$. 
            \If{$\llbracket Q_t[\overline{x},\overline{y}] \rrbracket_G = \emptyset$}
            \State Violated $\gets$ Violated $\cup$ $h_s[\overline{x}]$
            \EndIf
            \If {$\not\exists h_t[\overline{x},\overline{y}] \in \llbracket Q_t[\overline{x},\overline{y}] \rrbracket_G \models \phi_t$}
            \State Violated $\gets$ Violated $\cup$ $h_s[\overline{x}]$
            \EndIf
            \EndIf
            \EndFor
        \State \textbf{return} Violated \Comment{If the GGD $\sigma$ is validated then the Violated set is empty}
        \EndProcedure
    \end{algorithmic}
\end{algorithm}

Previous works proposed such algorithm in their own scenarios using a parallel algorithm~\cite{Fan2016}. Following the same strategy, we implement our algorithm using Spark framework\footnote{\url{https://spark.apache.org/}}, more specifically, using the capabilities of SparkSQL to query and handle large scale data. 

  \begin{table*}[t]
\centering
\small
\begin{tabular}{llll}
\hline
Dataset                       & Number of Nodes & Number of Edges & Number of GGDs \\ \hline
DBPedia Athlete               & 1.3M              & 1.7M              & 4 \\
LinkedMDB                     & 2.2M              & 11.2M       & 6       \\
LDBC SNB                     & 11M              & 66M          & 6    \\ \hline
\end{tabular}
\caption{Datasets used in our experiments}
\label{tab:datasets}
\end{table*}  

We used real-world datasets to verify the feasibility of using GGDs on real-world data and a synthtetic dataset to verify the impact of different aspects of GGDs on execution time.  
The LinkedMDB and the DBPedia datasets are real-world RDF~\cite{decker2000} datasets, to use it in our implementation, we transformed it into property graphs. 
For DBPedia, we extracted a subset of nodes and relationships that refer to sports athletes, events and teams. We named this subset as DBPedia Athlete.
The LDBC SNB (Social Network Benchmark) dataset\footnote{\url{https://github.com/ldbc/ldbc_snb_datagen_spark}} is a synthetic dataset that can be generated with different scale factors. The scale factor indicates the size of the dataset, the numbers in~\autoref{tab:datasets} are the total number of edges and nodes according to scale factor (SF)  = $3$.
See \autoref{tab:datasets} for details about the datasets.

We manually defined the GGDs according to the schema of each graph.
In the case of LinkedMDB and DBPedia Athlete, besides the schema information of when we transformed it into a property graph, we also used the results of the RDFind algorithm.
RDFind~\cite{Kruse2016}\footnote{\url{https://hpi.de/naumann/projects/repeatability/data-profiling/cind-discovery-on-rdf-data.html}} is an algorithm for finding conditional inclusion dependencies on RDF data. As mentioned in Section~\ref{sec:related} conditional inclusion dependencies are a special case of \tgds. 

\begin{figure}[t]
\centering
  \includegraphics[width=0.55\linewidth]{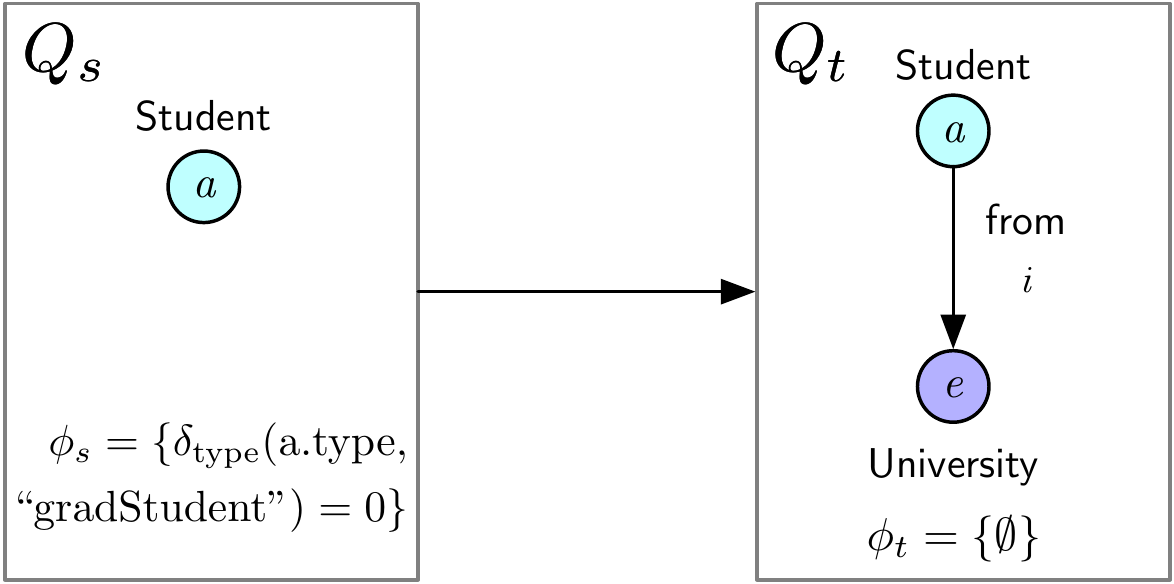}
  \caption{Including constraint information from a CIND in RDF into a GGD}
  \label{fig:CINDToGGD}
\end{figure}

Given the differences of the RDF and the property graph model, we cannot translate directly one CIND on RDF to a GGD on property graphs, instead, we included the main information given by the CINDs into the graph pattern of a GGD. Given an example CIND $(s, (p=rdf \wedge o = gradStudent)) \subseteq (s,p = underGradFrom)$, we show on \autoref{fig:CINDToGGD} how we used the information of this CIND in a GGD. 

    
For the LDBC dataset we manually defined a set of 6 different GGDs, in which 3 of them are fully validated and, 3 of them are highly violated. For the LinkedMDB and the DBPedia dataset, we defined a set of 6 and 4, respectively, of different GGDs, each taking into consideration the RDFind algorithm results. The set of GGDs defined for each dataset can be found on Appendix. We defined as our input, GGDs in which the target graph pattern has at least one common variable to the source graph pattern.

To implement the GGDs data inconsistencies algorithm, we used the G-Core language interpreter\footnote{\url{https://ldbcouncil.org/gcore-spark/}}~\cite{Angles2018} to query graph patterns using SparkSQL and implemented the algorithm in Scala using the Spark framework. The G-Core language is used for querying the graph patterns, the G-Core language interpreter is responsible for rewritting the G-Core queries into SparkSQL queries.
We use a relational representation of the property graphs in which each table represents a label.

Given this setup, we implemented two versions of the validation for data inconsistencies algorithm using SparkSQL\footnote{Our implementation is available at \url{https://github.com/smartdatalake/gcore-spark-ggd}}. We show both versions as a simplified set of commands in SparkSQL dataframe API in Algorithms~\ref{algo:leftanti} and \ref{algo:outer}.

\begin{algorithm}[htb]
    \caption{\textsc{LeftAnti}Join-Validation}
    \label{algo:leftanti}
    \begin{algorithmic} 
        \Procedure{DataInconsistency}{$ \sigma = Q_s[\overline{x}]\phi_s \rightarrow Q_t[\overline{x},\overline{y}]\phi_t \in \Sigma$} \Comment{This procedure is repeated for each GGD in the set $\Sigma$}
            \State source\_df  $\gets \{h(Q_s[\overline{x})\}$  \Comment{Search for matches of the source graph pattern}
            \State source\_df.filter($\phi_s$) \Comment{Checks which matches of the source graph pattern holds the constraints in $\phi_s$}
            \State target\_df $\gets \{h(Q_t[\overline{x},\overline{y}]\}$
            \State target\_df.filter($\phi_{t1}$) \Comment{Set of matches of the target graph pattern that holds the constraints in $\phi_t$ that are exclusively constraints of variables that are present only in the target graph pattern}
            \State source\_target\_df $\gets$ source\_df INNER JOIN target\_df ON $\overline{x}$
            \State source\_target\_df.filter($\phi_{t2}$) \Comment{All validated matches}
            \State violated\_df $\gets$ source\_df LEFT ANTI JOIN source\_target\_df ON $\overline{x}$
            \State {return} violated
        \EndProcedure
    \end{algorithmic}
\end{algorithm}

\begin{algorithm}[htb]
    \caption{\textsc{LeftOuter}Join-Validation}
    \label{algo:outer}
    \begin{algorithmic}
        \Procedure{DataInconsistency}{$ \sigma = Q_s[\overline{x}]\phi_s \rightarrow Q_t[\overline{x},\overline{y}]\phi_t \in \Sigma$} \Comment{This procedure is repeated for each GGD in the set $\Sigma$}
            \State source\_df  $\gets \{h(Q_s[\overline{x})\}$  \Comment{Search for matches of the source graph pattern}
            \State target\_df $\gets \{h(Q_t[\overline{x},\overline{y}]\}$
            \State source\_target\_df $\gets$ source\_df LEFT OUTER JOIN target\_df ON $\overline{x}$
            \State source\_target\_df.filter($\phi_s$) \Comment{All matches of the source that should be validated}
            \State violated\_df $\gets$ source\_target\_df.filter($\overline{\phi_t} || (\not\exists \overline{y}$))
            \State {return} violated
        \EndProcedure
    \end{algorithmic}
\end{algorithm}

In the first version, we used ``left anti joins'' to check for data inconsistencies and in the second version we used ``left outer joins''. This choice of operators was based on the works on validation over \tgds of the literature in the scenario of validating schema mapping~\cite{Alexe2011,Bonifati2008}. These works use the EXISTS operator in SQL to verify if the schema mapping can be validated, in our case we translate it to left-anti and left-outer joins since in here we are dealing with (possibly) different graph patterns.
While there is room for optimization and improvement on the implementation, the goal is to show how GGDs are feasible even when using an available query engine such as SparkSQL.

For checking the differential constraints, we applied two strategies: (1) if the differential constraint compares two attributes (properties) of two different nodes or edges which are connected in the graph pattern,  or, the differential constraint compares an attribute (property) to a constant value, then we check if the differential constraint holds on each graph pattern match (linear scan of the graph pattern match table);
(2) if the differential constraints compare two attributes (properties) of two different nodes of edges that are not connected. To avoid a Cartesian product, we include similarity search and joins operators to speed up the process. In the implementation, we used Dima's system~\cite{Ji2019} operators and integrated them into SparkSQL for both Jaccard and Edit distance. In this case, we use the similarity join between the matches of each disconnected part of the graph pattern.

\begin{table*}[ht!]
\centering
\small
\begin{tabular}{llllll}
Dataset & GGD & Source matches  & Violated matches & \textsc{LeftAnti} & \textsc{LeftOuter} \\ \hline
LDBC &1 & 3182834 & 3078209 & 5.66 & 6.38 \\
LDBC & 2 & 270602 &	0 & 0.48 & 0.53 \\
LDBC &3 & 3835041 & 0 & 3.39 & 4.6 \\
LDBC &4 & 634081 & 0 & 1.21 & 4.93 \\
LDBC & 5 & 10477317 &9387084 & 73.41 & 205.73\\
LDBC &6 & 27133 &	26265 & 1.76 & 3.33 \\
\hline
LinkedMDB &1 & 21966 & 3603 & 3.04 & 3.48 \\
LinkedMDB &2 & 15220 &	0 &	4.11 & 4.54 \\
LinkedMDB &3 & 98816 &	98812 & 2.94 & 3.31 \\
LinkedMDB &4 & 71734 &	71547 &	2.52 & 2.84 \\
LinkedMDB &5 & 1613873 & 329718 & 6.72 & 7.41 \\
LinkedMDB &6 & 196367 & 194841 & 4.18 & 4.48 \\
\hline
DBPedia& 1 & 9495 & 9495	& 7.61 & 7.59 \\
DBPedia& 2 & 2388 & 1089 & 5.42 & 5.41 \\ 
DBPedia& 3 & 17 & 17 & 4.9 & 4.87 \\
DBPedia& 4 & 389 & 386 & 4.63 & 4.55 \\
\hline
\end{tabular}
\caption{Number of source matches to be validated and number of violated matches according to the GGDs of the tested dataset. The execution time shown in \textsc{LeftAnti} and \textsc{LeftOuter} are in minutes.}
\label{tab:selectivityAll}
\end{table*}

\subsection{Impact of the selectivity of each GGD}
\label{subsec:selectivity}

In this section, we analyse the selectivity of each GGD, this means how the number of matches of the source and number violations can affect the execution time.
To do so, we executed each one of the GGDs of each one of the datasets separately in both of the implementations we presented earlier. Table \ref{tab:selectivityAll} shows the results of this experiment.

From these results, we can observe that overall the greater the number of source matches and the smaller the number of violations the more time it takes to execute the validation algorithm. 
It also depends not only on the number of matches but also on the graph pattern in itself. This is related to conjunctive query evaluation mentioned in Section~\ref{sec:validation} (see also~\cite{Bonifati2008} for a study of the main aspects that can affect the execution time of a SPARQL~\cite{Perez2006} graph query). 

We can also observe that in the case of DBPedia Athlete, which is the smallest graph we used in these experiments, the difference in execution time between the validation using the \textsc{LeftAnti} algorithm and the \textsc{LeftOuter} algorithm are very similar, being the \textsc{LeftOuter} algorithm faster than than the \textsc{LeftAnti}. However, if we compare the execution time of both algorithms in the LinkedMDB dataset or the LDBC dataset, in both cases, the left outer algorithm is slower. This difference is more pronounced when the number of source graph patterns matched are higher (see GGD number 5 of LDBC and LinkedMDB in \autoref{tab:selectivityAll}).

\subsection{Impact of the differential constraints threshold}
\label{subsec:relation}

\begin{figure*}[htbp]
\centering
 \includegraphics[width=0.6\linewidth]{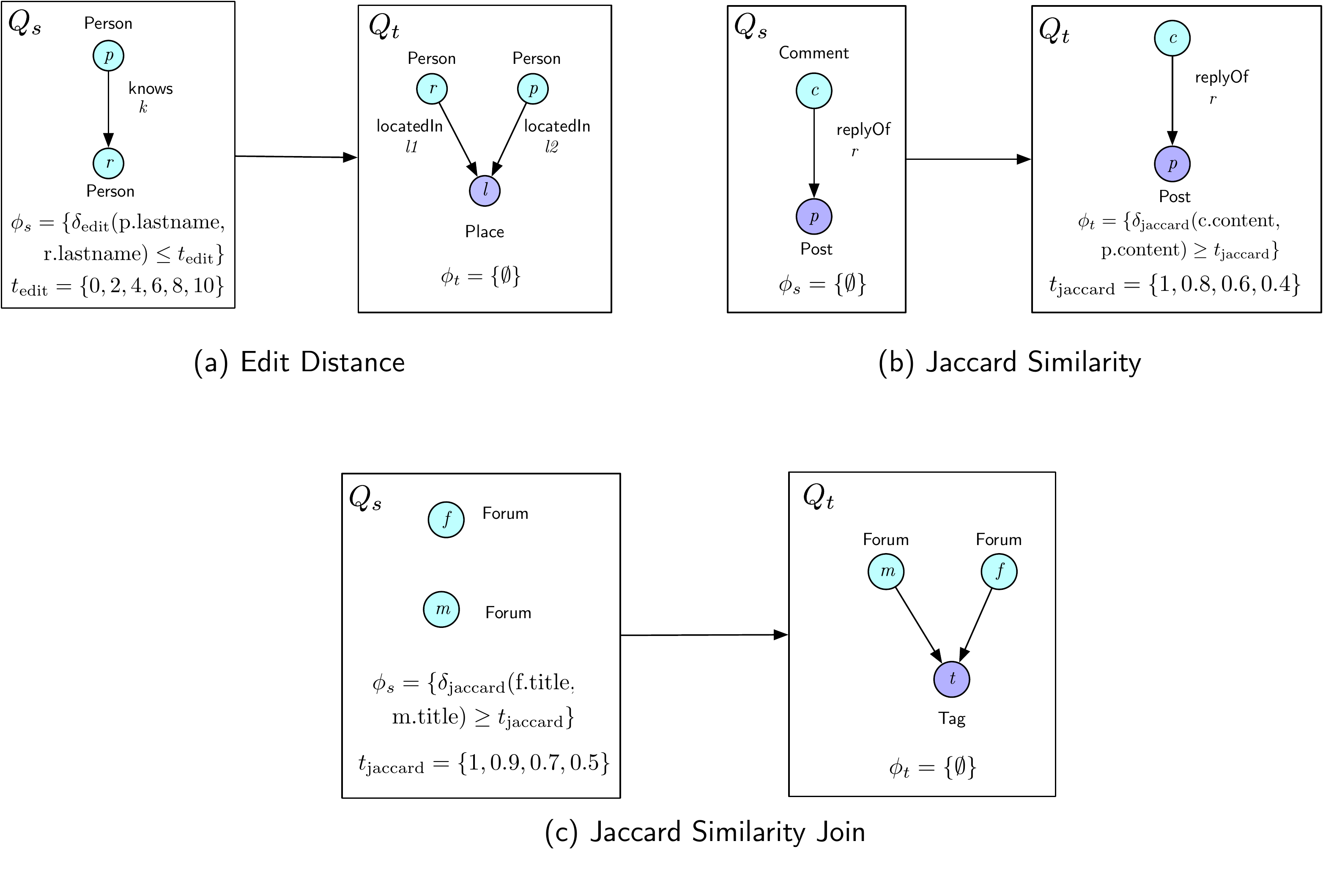}
  \caption{Tested GGDs according to the differential constraint threshold}
  \label{fig:GGDsTest}
\end{figure*}

\begin{figure}[htbp]
 \includegraphics[width=\linewidth]{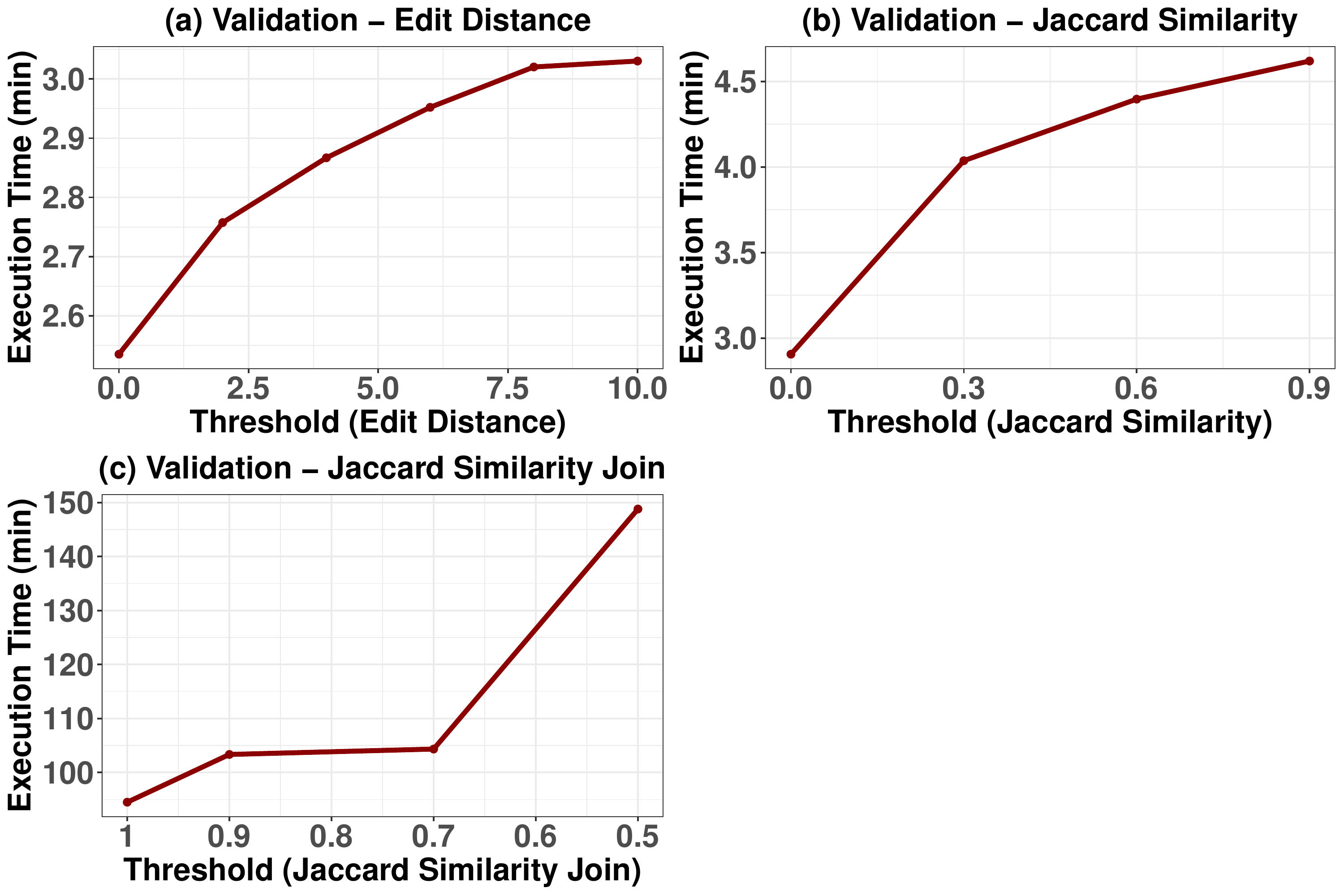}
  \caption{Execution time of the GGDs in \autoref{fig:GGDsTest} according to the differential constraint threshold (Left Anti Join)}
  \label{fig:GGDsTestExecTime}
\end{figure}

Next, we used three different GGDs to test their behavior according to the increase of the threshold in the differential constraints. The three different GGDs tested are in \autoref{fig:GGDsTest} and its respective execution time are presented in \autoref{fig:GGDsTestExecTime}.
The GGDs (a) and (b) are GGDs in which there is a connection in the graph pattern between the nodes/edges. In this case, the constraint checking procedure is to check match by match of the graph pattern if the differential constraint holds for that match. Observe in (a) and (b) the execution time results according to the increase of the threshold for both of these GGDs. 

A special case of GGD is the GGDs represented by (c) in which the only connection between the nodes in the graph pattern is the differential constraint. In this case, when searching for the source graph pattern matches alone would result in the execution of a cartesian product. To avoid this expensive operation, we use a similarity join operator to search for the source graph patterns. We first match each disconnected part of the graph pattern separately and then use the similarity join operator to find the matches of the source side.

Observe that the difference in execution time according to the threshold of the cases (a) and (b) are very small compared to (c). This happens because, since the constraint checking procedure is done match per match, the number of times the constraint will be checked is the same, as the number of matches of the graph pattern, independently of the threshold. This is aligned with the complexity analysis of the validation algorithm in Section~\ref{sec:validation} in which the graph pattern is the most expensive "part" of matching the source or the target. In this case, the increase in execution time is because of the number of source matches that need to be validated, as the threshold increases the number of source matches to be validated also increases. As an example, \autoref{tab:editConstraints} shows the number of source matches to be validated according to GGD (a). 

However, when there is a need for the similarity join operator, the increase in execution time is more accentuated by the increase of the threshold because, besides the consequent increase in the number of sources to be validated, the cost of performing a similarity join increases as well (see~\cite{Jiang2014} for more details on string similarity join algorithms).

\begin{table}[t]
\centering
\small
\begin{tabular}{lll}
Source matches  & Violated matches & Threshold (Edit distance) \\ \hline
27133	& 26265 & 0 \\
50018	& 48952 & 2 \\
187709	& 185247 & 4 \\
493646	& 488217 & 6 \\
650125	& 643051 & 8 \\
688305	& 680805 & 10 \\
\hline
\end{tabular}
\caption{Number of Source Matches to be Validated and Number of Violated Matches according to GGD (a) of \autoref{fig:GGDsTest} for LDBC Dataset (scale factor = 3)}
\label{tab:editConstraints}
\end{table}

\subsection{Impact of the size of the data}
\label{subsec:scalability}

To analyze the scalability of the validation algorithm we analyzed how the same set of GGDs affects time execution according to the increasing size of data. We generated the LDBC Social Network dataset with different scale factors to simulate the increasing size of data. The Table~\ref{tab:ldbcScalability} contains information about the approximate number of nodes and edges according to the different scale factors.


\begin{table}[t]
\centering
\small
\begin{tabular}{lll}
Scale Factor      & Number of Nodes & Number of Edges \\ \hline
0.1                & 500.000              & 2M              \\
0.3                 & 1M              & 6M              \\
1.0                 & 4M              & 22M              \\ 
3.0                  & 11M              & 66M              \\
\hline
\end{tabular}
\caption{LDBC Size according to different Scale Factors}
\label{tab:ldbcScalability}
\end{table}

\begin{figure}[t]
  \centering
   \includegraphics[width=0.71\linewidth]{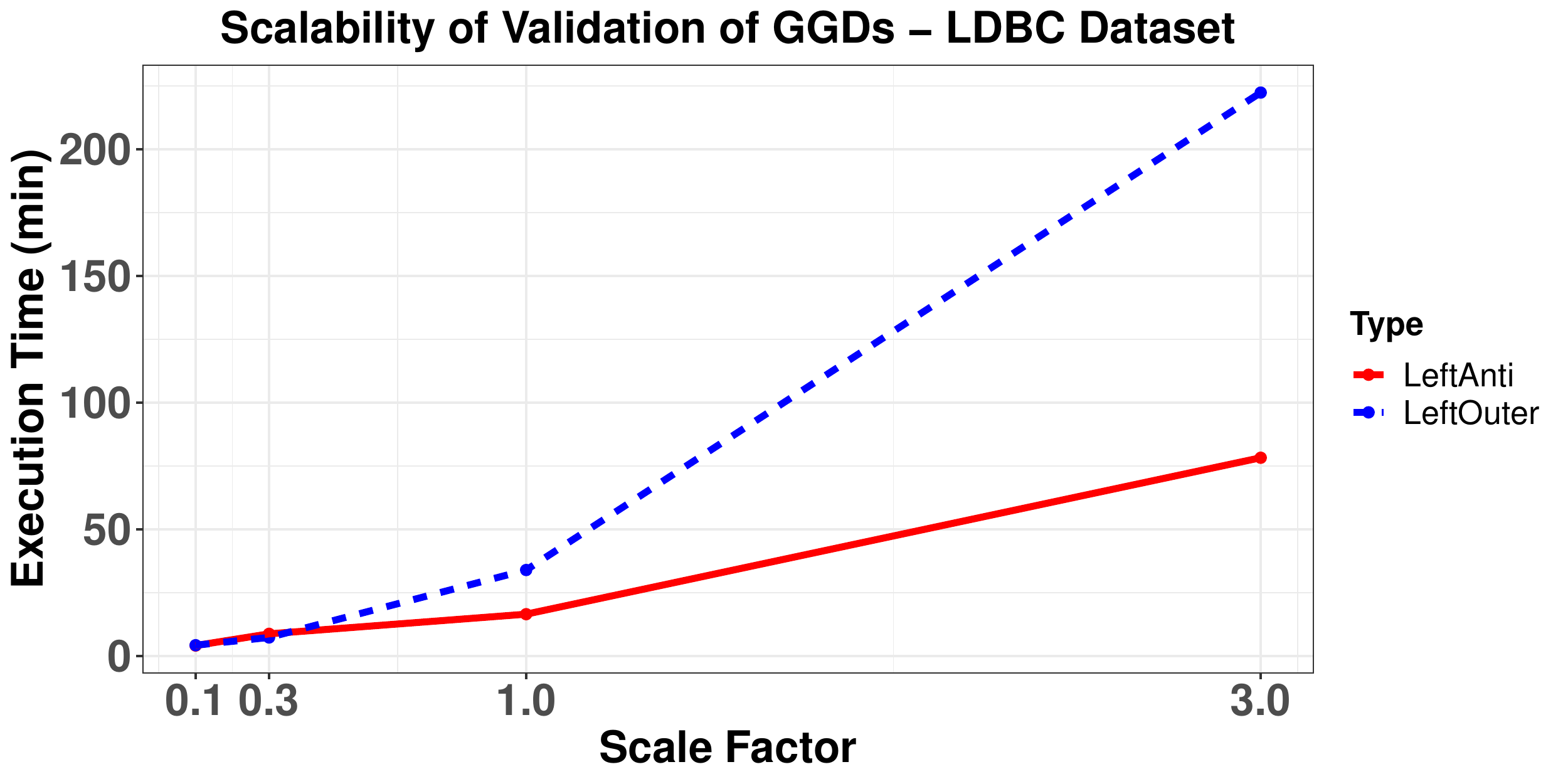}
 \caption{Scalability of the validation according to the Scale Factor of LDBC}
  \label{fig:scalabilityResults}
\end{figure}

The results in \autoref{fig:scalabilityResults} shows that as the data size grows the execution time also grows, as expected. We can see in special an substantial increase in the execution time for both versions between the scale factor $0.3$ and $1$ as both number of nodes and edges have almost quadruplicated. 

Lastly, we run the full validation on the set of GGDs in each one of the datasets. The execution time for all of the datasets are in Table~\ref{tab:fullValidationTime}. In all of the datasets, the full validation of the set of GGDs was completed in a feasible time for both versions of implementation. 
For DBPedia Athlete and LinkedMDB, the execution time was similar, while for LDBC the execution time for the Left Outer algorithm was significantly worst, following the results reported in Subsection~\ref{subsec:scalability}.
Given the expressivity power of GGDs and the ability to use user-defined differential constraints a small set of GGDs, such as the ones tested in these experiments, depending on the data, can be enough to represent the interesting constraints for different practical applications.


\begin{table*}[t]
\centering
\small
\begin{tabular}{llll}
Dataset & Number of GGDs & LeftAnti (in min.) & Outer (in min.) \\ \hline
DBPedia Athlete & 4 & 22.48 & 22.46\\
LinkedMDB & 6 & 25.75 & 25.77 \\
LDBC & 6 & 78.25 & 222.37 \\
\hline
\end{tabular}
\caption{Execution time for full validation of the GGDs}
\label{tab:fullValidationTime}
\end{table*}

\subsection{Experimental Results Conclusion}

In the experiments presented in this section, we showed that even though the validation algorithm has high complexity (Section~\ref{sec:validation}), we showed that using GGDs for identifying data inconsistency is feasible. Thus, the algorithm can be easily implemented using an available platform/framework.

There are, nevertheless, a lot of space for optimizing the algorithm presented, for example, new techniques for graph pattern matching, similarity searching (for the differential constraints), and the use of other types of frameworks or architectures that are more suitable to the problem (improving the Spark framework setup instead of running in a single machine, for example).
To understand the behavior and which aspects of GGDs can affect the execution time (and should later be taken into consideration in algorithm optimizations), we evaluated three main aspects: (1) how the number of source graph pattern matches affects time, (2) how the threshold in the differential constraint affects time, and (3) how the algorithm scales (according to this implementation).   

From the results, we could observe that the higher the number of source matches to be evaluated the higher the execution time. This also correlates to the differential constraints threshold as the bigger the threshold in the differential the higher the number of source matches to evaluate. However, it also depends on the cost of the source graph pattern evaluation, i.e., a GGD that has a more costly source graph pattern takes more time to execute compared to a less costly one even when the number the of source matches and violated are the same. These two aspects (1) number of source matches and (2) cost of the graph pattern evaluation should be taken into consideration when proposing optimization techniques. 
We also verified that for the same set of GGDs, the algorithm that uses left-anti joins scales better than the one that uses left-outer joins, giving some indication on which operators/strategy fits better in terms of implementation for a better scalable algorithm.
In future work, we plan take these results into consideration and propose an optimized version of the evaluated algorithm.  


\section{Conclusion and Future Work} 

Motivated by practical applications in graph data management,
we studied three main reasoning problems for a new class of graph dependencies called Graph Generating Dependencies (GGDs). GGDs are inspired by the tuple-generating dependencies and the equality-generating dependencies from relational data, where constraint satisfaction can generate new vertices and edges. 

In this paper, we proposed algorithms for solving three fundamental reasoning problems of GGDs, satisfiability, implication, and validation. To prove the complexity of such algorithms we propose a Chase procedure for GGDs based on the standard Chase for \tgds and GEDs. 
Even though our results show that the reasoning problems of GGDs have high complexity, in our experimental results we verify how GGDs are feasible to be used in practice to identify data inconsistencies. In our experimental results, we also analysed the impact of the size of the data and the differential constraints on a set of GGDs.
In future work, we plan to study the discovery of GGDs and their applications in graph data profiling and cleaning.

\smallskip
    \noindent {\bf Acknowledgments.} This project has received funding from the European Union's Horizon 2020 research and innovation programme under grant agreement No 825041.

\bibliography{mybibfile}

\appendix
\let\cleardoublepage\clearpage

\section{Input GGDs}
\label{app:inputggds}

This appendix contains the input GGDs that were manually defined for each one of the datasets presented in the Section~\ref{sec:experiments}. The GGDs follows the same order as presented in the results tables of Subsection~\ref{subsec:selectivity}.

\let\cleardoublepage\clearpage

\begin{figure*}
  \centering
  \includegraphics*[height=0.95\textheight]{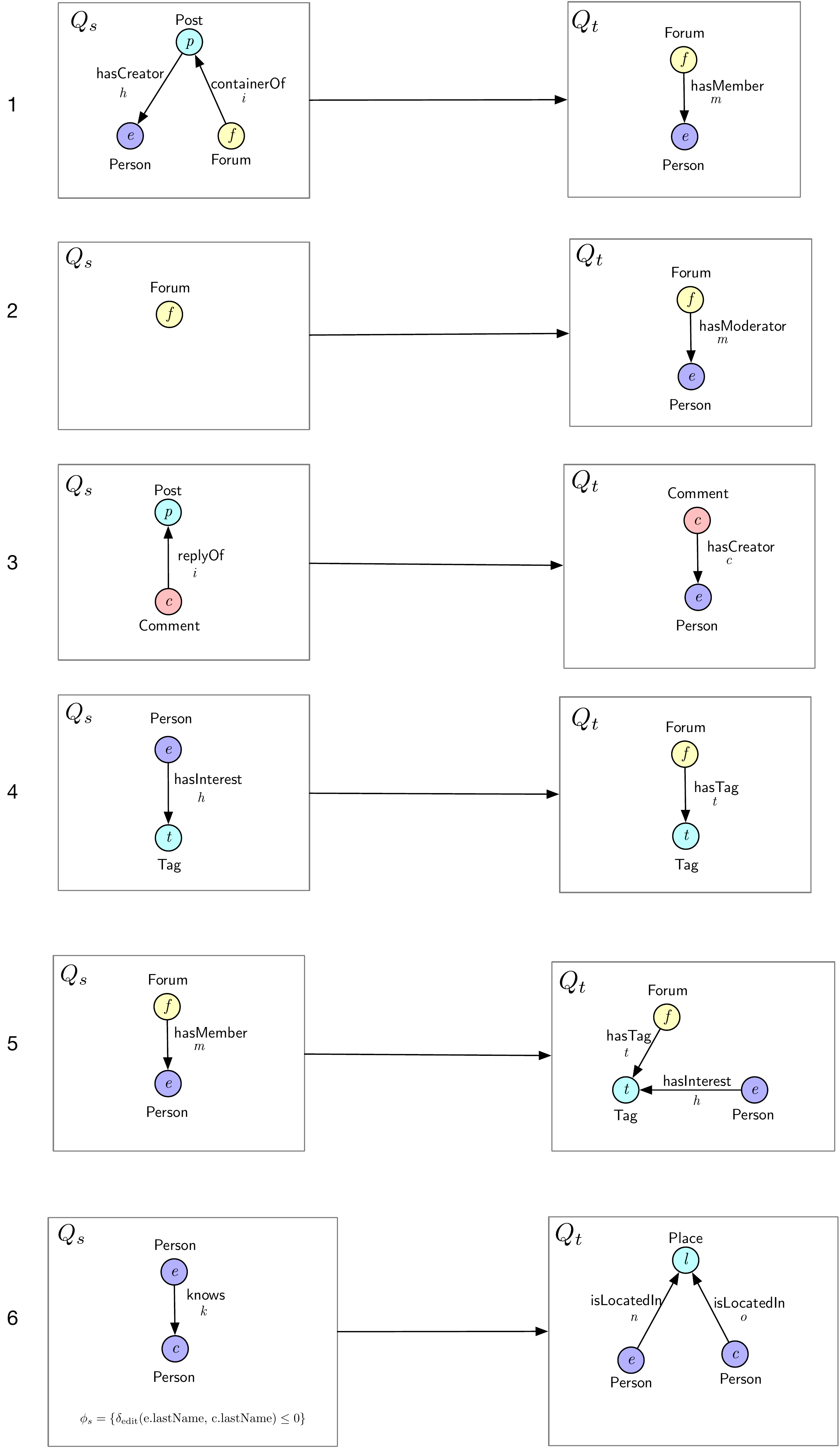}
  \label{fig:inputGGDsLDBC}
  \caption{Input GGDs: LDBC Dataset}
\end{figure*}

\let\cleardoublepage\clearpage

\begin{figure*}
  \centering
  \includegraphics*[height=0.95\textheight]{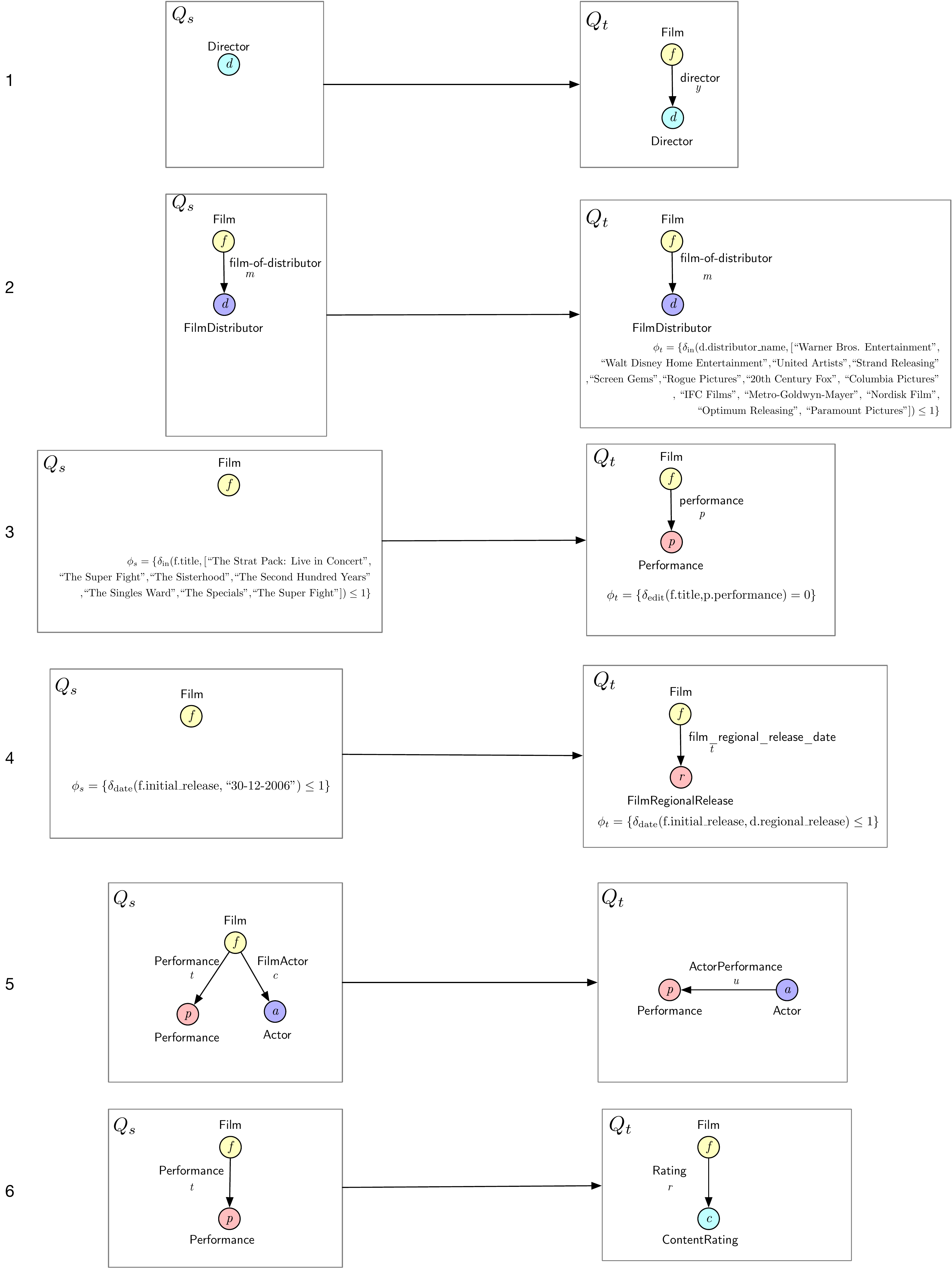}
  \label{fig:inputGGDsLinkedMDB}
  \caption{Input GGDs: LinkedMDB Dataset}
\end{figure*}

\let\cleardoublepage\clearpage

\begin{figure*}
  \centering
  \includegraphics*[height=0.95\textheight]{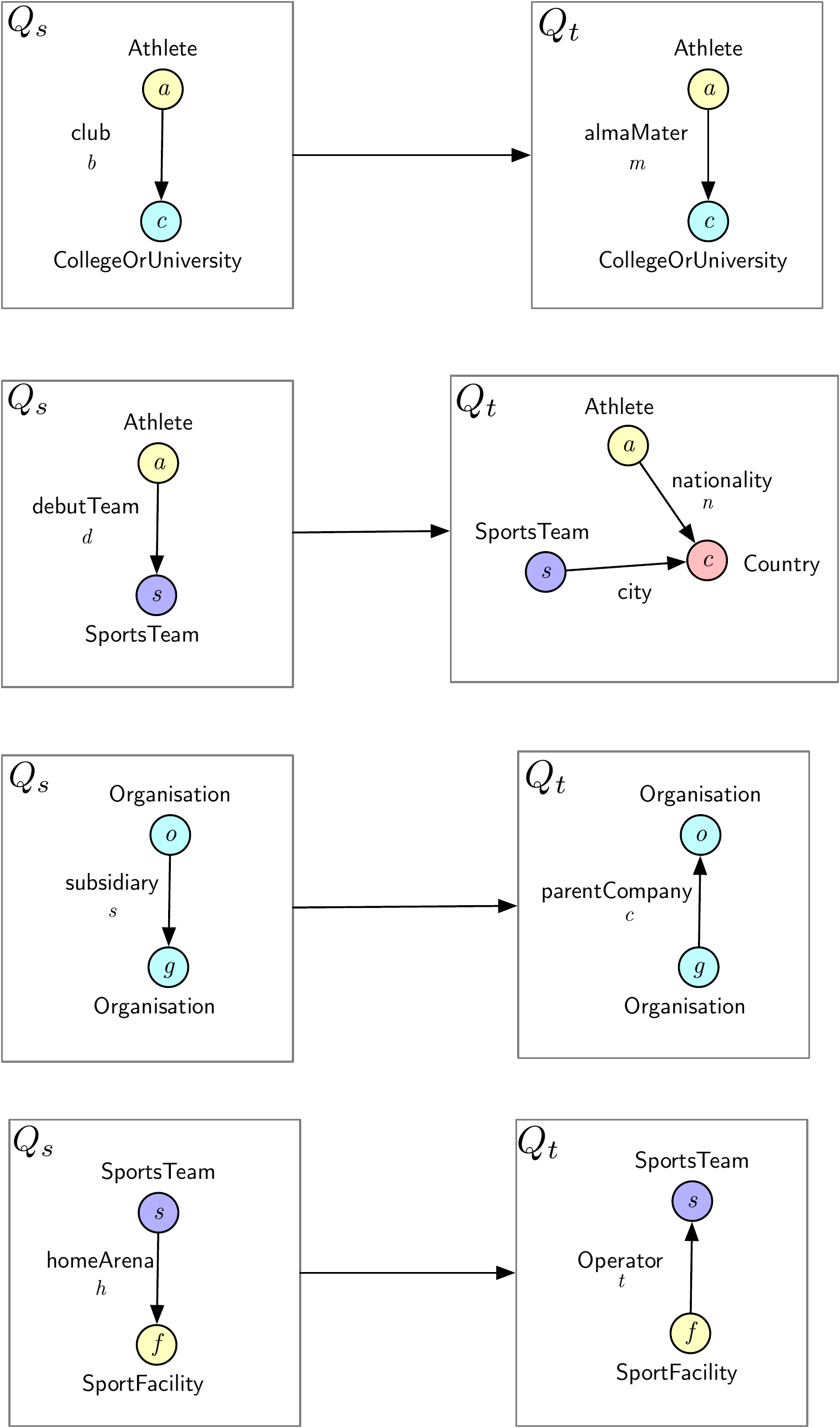}
  \label{fig:inputGGDsDBPedia}
    \caption{Input GGDs: DBPedia Athlete}
\end{figure*}

\end{document}